\documentclass[12pt]{article}

\usepackage{axodraw}

\makeatletter

\def\paragraph{\@startsection{paragraph}{4}{\z@}{+2.00ex plus
 +1ex minus +.2ex}{1.5ex plus .2ex}{\it\normalsize}}

\def\section{\@startsection {section}{1}{\z@}{+3.0ex plus +1ex minus
  +.2ex}{2.3ex plus .2ex}{\normalsize\bf\boldmath}}
\def\subsection{\@startsection{subsection}{2}{\z@}{+2.5ex plus +1ex
minus +.2ex}{1.5ex plus .2ex}{\normalsize\bf\boldmath}}
\def\subsubsection{\@startsection{subsubsection}{3}{\z@}{+3.25ex plus
 +1ex minus +.2ex}{1.5ex plus .2ex}{\normalsize\it}}

\expandafter\ifx\csname mathrm\endcsname\relax\def\mathrm#1{{\rm #1}}\fi

\renewcommand{\theequation}{\thesection.\arabic{equation}}
\newcounter{saveeqn}

\@addtoreset{equation}{section}

\newcount\@tempcntc
\def\@citex[#1]#2{\if@filesw\immediate\write\@auxout{\string\citation{#2}}\fi
  \@tempcnta\z@\@tempcntb\m@ne\def\@citea{}\@cite{\@for\@citeb:=#2\do
    {\@ifundefined
       {b@\@citeb}{\@citeo\@tempcntb\m@ne\@citea
        \def\@citea{,\penalty\@m\ }{\bf ?}\@warning
       {Citation `\@citeb' on page \thepage \space undefined}}%
    {\setbox\z@\hbox{\global\@tempcntc0\csname
b@\@citeb\endcsname\relax}%
     \ifnum\@tempcntc=\z@ \@citeo\@tempcntb\m@ne
       \@citea\def\@citea{,\penalty\@m}
       \hbox{\csname b@\@citeb\endcsname}%
     \else
      \advance\@tempcntb\@ne
      \ifnum\@tempcntb=\@tempcntc
      \else\advance\@tempcntb\m@ne\@citeo
      \@tempcnta\@tempcntc\@tempcntb\@tempcntc\fi\fi}}\@citeo}{#1}}

\def\@citeo{\ifnum\@tempcnta>\@tempcntb\else\@citea
  \def\@citea{,\penalty\@m}%
  \ifnum\@tempcnta=\@tempcntb\the\@tempcnta\else
   {\advance\@tempcnta\@ne\ifnum\@tempcnta=\@tempcntb \else
\def\@citea{--}\fi
    \advance\@tempcnta\m@ne\the\@tempcnta\@citea\the\@tempcntb}\fi\fi}

\newcommand{\ks}{k\hspace{-0.52em}/\hspace{0.1em}}
\newcommand{\lslash}{l\hspace{-0.42em}/\hspace{0.1em}}
\newcommand{\ps}{p\hspace{-0.42em}/}
\newcommand{\qs}{q\hspace{-0.48em}/}

\def\nl{\nonumber\\}

\def\nln{\nonumber\\*[-1ex]\phantom{\fbox{\rule{0em}{2ex}}}}

\def\asymp#1%
{\mathrel{\raisebox{-.4em}{$\widetilde{\scriptstyle #1}$}}}

\def\Nequal#1%
{\mathrel{\raisebox{-.5em}{$\stackrel{=}{\scriptstyle\rm#1}$}}}
\newcommand{\dsl}[1]{\not \hspace{-0.7mm}#1}
\def\dsl{\mathpalette\make@slash}
\def\make@slash#1#2{\setbox\z@\hbox{$#1#2$}%
  \hbox to 0pt{\hss$#1/$\hss\kern-\wd0}\box0}

\def\beq{\begin{equation}}
\def\eeq{\end{equation}}
\def\beqar{\begin{eqnarray}}
\def\eeqar{\end{eqnarray}}
\def\barr#1{\begin{array}{#1}}
\def\earr{\end{array}}
\def\bfi{\begin{figure}}
\def\efi{\end{figure}}
\def\btab{\begin{table}}
\def\etab{\end{table}}
\def\bce{\begin{center}}
\def\ece{\end{center}}
\def\nn{\nonumber}

\def\text{\textstyle}

\def\de{\delta}
\def\teps{\varepsilon}
\def\veps{\epsilon}
\def\la{\lambda}
\def\si{\sigma}

\def\refeq#1{\mbox{(\ref{#1})}}

\def\refse#1{\mbox{Sect.~\ref{#1}}}
\def\refses#1{\mbox{Sects.~\ref{#1}}}
\def\refapp#1{\mbox{App.~\ref{#1}}}

\def\citere#1{\mbox{Ref.~\cite{#1}}}
\def\citeres#1{\mbox{Refs.~\cite{#1}}}

\def\solid{\raise.9mm\hbox{\protect\rule{1.1cm}{.2mm}}}
\def\dash{\raise.9mm\hbox{\protect\rule{2mm}{.2mm}}\hspace*{1mm}}

\newcommand{\TeV}{\unskip\,\mathrm{TeV}}

\def\mathswitchr#1{\relax\ifmmode{\mathrm{#1}}\else$\mathrm{#1}$\fi}

\newcommand{\PW}{\mathswitchr W}
\newcommand{\PZ}{\mathswitchr Z}

\newcommand{\PA}{\mathswitchr A}

\newcommand{\PH}{\mathswitchr H}

\newcommand{\Pt}{\mathswitchr t}

\newcommand{\Pep}{\mathswitchr {e^+}}
\newcommand{\Pem}{\mathswitchr {e^-}}

\newcommand{\PWpm}{\mathswitchr {W^\pm}}

\def\mathswitch#1{\relax\ifmmode#1\else$#1$\fi}

\newcommand{\MW}{\mathswitch {M_\PW}}

\newcommand{\MZ}{\mathswitch {M_\PZ}}
\newcommand{\MH}{\mathswitch {M_\PH}}

\newcommand{\Mt}{\mathswitch {m_\Pt}}

\newcommand{\scrs}{\scriptscriptstyle}
\newcommand{\sw}{\mathswitch {s_{\scrs\PW}}}
\newcommand{\cw}{\mathswitch {c_{\scrs\PW}}}
\newcommand{\swbare}{\mathswitch {s_{{\scrs\PW},0}}}
\newcommand{\cwbare}{\mathswitch {c_{{\scrs\PW},0}}}

\newcommand{\muD}{\mu_0}
\newcommand{\muR}{\mu_\mathrm{w}}
\newcommand{\mue}{\mu_e}

\newcommand{\betacoeff}[1]{b_{#1}}

\newcommand{\ctwo}{C_{\si}}

\newcommand{\vev}{{\bf v}}

\def\ie{i.e.\ }

\newcommand{\etal}{{\it et al.}}

\newcommand{\order}{\mathcal{O}}

\newcommand{\NLLA}{\stackrel{\mathrm{NLL}}{=}}

\newcommand{\Tr}{\mathrm{Tr}}

\newcommand{\SU}{\mathrm{SU}}

\newcommand{\SUtwo}{\mathrm{SU}(2)}
\newcommand{\Uone}{\mathrm{U}(1)}

\newcommand{\ri}{\mathrm{i}}

\newcommand{\rd}{{\mathrm{d}}}

\newcommand{\elm}{\mathrm{em}}

\newcommand{\sew}{\mathrm{sew}}

\newcommand{\QED}{{\mathrm{QED}}}

\newcommand{\deone}{\de^{\Uone}}
\newcommand{\detwo}{\de^{\SUtwo}}

\def\draftdate{\relax}
\def\mda{\relax}
\def\mua{\relax}
\def\mla{\relax}
\def\draft{
\def\thtystars{******************************}
\def\sixtystars{\thtystars\thtystars}
\typeout{}
\typeout{\sixtystars**}
\typeout{* Draft mode!
         For final version remove \protect\draft\space in source file *}
\typeout{\sixtystars**}
\typeout{}
\def\draftdate{\today}
\def\mua{\marginpar[\boldmath\hfil$\uparrow$]%
                   {\boldmath$\uparrow$\hfil}%
                    \typeout{marginpar: $\uparrow$}\ignorespaces}
\def\mda{\marginpar[\boldmath\hfil$\downarrow$]%
                   {\boldmath$\downarrow$\hfil}%
                    \typeout{marginpar: $\downarrow$}\ignorespaces}
\def\mla{\marginpar[\boldmath\hfil$\rightarrow$]%
                   {\boldmath$\leftarrow $\hfil}%
                    \typeout{marginpar: $\leftrightarrow$}\ignorespaces}
\def\Mua{\marginpar[\boldmath\hfil$\Uparrow$]%
                   {\boldmath$\Uparrow$\hfil}%
                    \typeout{marginpar: $\Uparrow$}\ignorespaces}
\def\Mda{\marginpar[\boldmath\hfil$\Downarrow$]%
                   {\boldmath$\Downarrow$\hfil}%
                    \typeout{marginpar: $\Downarrow$}\ignorespaces}
\def\Mla{\marginpar[\boldmath\hfil$\Rightarrow$]%
                   {\boldmath$\Leftarrow $\hfil}%
                    \typeout{marginpar: $\Leftrightarrow$}\ignorespaces}

\overfullrule 5pt
\oddsidemargin -15mm
\marginparwidth 29mm
}

\makeatletter

\def\eqnarray{\stepcounter{equation}\let\@currentlabel=\theequation
\global\@eqnswtrue
\global\@eqcnt\z@\tabskip\@centering\let\\=\@eqncr
$$\halign to \displaywidth\bgroup\hskip\@centering
  $\displaystyle\tabskip\z@{##}$\@eqnsel&\global\@eqcnt\@ne
  \hskip 2\arraycolsep \hfil${##}$\hfil
  &\global\@eqcnt\tw@ \hskip 2\arraycolsep $\displaystyle\tabskip\z@{##}$\hfil
   \tabskip\@centering&\llap{##}\tabskip\z@\cr}

\makeatother

\begin{document}

\newcommand{\leg}[1]{}
\newcommand{\diaggeneric}{
\begin{picture}(120.,104.)(-28.,-52.)
\Gluon(-53.6656,0.)(0.,0.){3.2}{5}
\ArrowLine(0.,0.)(24.,12.)
\ArrowLine(24.,12.)(48.,24.)
\ArrowLine(24.,-12.)(0.,0.)
\ArrowLine(48.,-24.)(24.,-12.)
\GCirc(4.47214,0.){17}{0.8}
\Text(84.,42.)[l]{$\leg{1}$}
\Text(84.,-42.)[l]{$\leg{2}$}
\Text(27.668,14.6453)[br]{}
\Text(28.,-14.)[tr]{}
\end{picture}
}

\newcommand{\diagone}{
\begin{picture}(120.,104.)(-28.,-52.)
\Gluon(-35.7771,0.)(0.,0.){3.2}{4}
\Vertex(0.,0.){2}
\ArrowLine(0.,0.)(48.,24.)
\ArrowLine(48.,24.)(80.,40.)
\ArrowLine(48.,-24.)(0.,0.)
\ArrowLine(80.,-40.)(48.,-24.)
\Photon(48.,24.)(48.,-24.){2.4}{3.5}
\Vertex(48.,24.){2}
\Vertex(48.,-24.){2}
\Text(84.,42.)[l]{$\leg{1}$}
\Text(84.,-42.)[l]{$\leg{2}$}
\Text(53.6656,0.)[l]{$\scriptstyle{a_1}$}
\Text(27.668,14.6453)[br]{}
\Text(28.,-14.)[tr]{}
\end{picture}
}

\newcommand{\diagI}{
\begin{picture}(120.,104.)(-28.,-52.)
\Gluon(-35.7771,0.)(0.,0.){3.2}{4}
\Vertex(0.,0.){2}
\ArrowLine(0.,0.)(28.,14.)
\ArrowLine(28.,14.)(56.,28.)
\ArrowLine(56.,28.)(80.,40.)
\ArrowLine(28.,-14.)(0.,0.)
\ArrowLine(56.,-28.)(28.,-14.)
\ArrowLine(80.,-40.)(56.,-28.)
\Photon(56.,28.)(56.,-28.){2.4}{3.5}
\Photon(28.,14.)(28.,-14.){2.4}{2}
\Vertex(56.,28.){2}
\Vertex(56.,-28.){2}
\Vertex(28.,14.){2}
\Vertex(28.,-14.){2}
\Text(84.,42.)[l]{$\leg{1}$}
\Text(84.,-42.)[l]{$\leg{2}$}
\Text(62.6099,0.)[l]{$\scriptstyle{a_1}$}
\Text(31.305,0.)[l]{$\scriptstyle{a_2}$}
\Text(41.5019,21.9679)[br]{}
\Text(42.,-21.)[tr]{}
\Text(19.515,10.9162)[br]{}
\Text(20.,-10.)[tr]{}
\end{picture}
}

\newcommand{\diagII}{
\begin{picture}(120.,104.)(-28.,-52.)
\Gluon(-35.7771,0.)(0.,0.){3.2}{4}
\Vertex(0.,0.){2}
\ArrowLine(0.,0.)(28.,14.)
\ArrowLine(28.,14.)(56.,28.)
\ArrowLine(56.,28.)(80.,40.)
\ArrowLine(28.,-14.)(0.,0.)
\ArrowLine(56.,-28.)(28.,-14.)
\ArrowLine(80.,-40.)(56.,-28.)
\Photon(56.,28.)(28.,-14.){-2.4}{3}
\Photon(56.,-28.)(28.,14.){-2.4}{3}
\Vertex(56.,28.){2}
\Vertex(56.,-28.){2}
\Vertex(28.,14.){2}
\Vertex(28.,-14.){2}
\Text(84.,42.)[l]{$\leg{1}$}
\Text(84.,-42.)[l]{$\leg{2}$}
\Text(48.3499,9.07117)[l]{$\scriptstyle{a_1}$}
\Text(48.3499,-9.07117)[l]{$\scriptstyle{a_2}$}
\Text(41.5019,21.9679)[br]{}
\Text(42.,-21.)[tr]{}
\Text(19.515,10.9162)[br]{}
\Text(20.,-10.)[tr]{}
\end{picture}
}

\newcommand{\diagIII}{
\begin{picture}(120.,104.)(-28.,-52.)
\Gluon(-35.7771,0.)(0.,0.){3.2}{4}
\Vertex(0.,0.){2}
\ArrowLine(0.,0.)(24.,12.)
\ArrowLine(24.,12.)(56.,28.)
\ArrowLine(56.,28.)(80.,40.)
\ArrowLine(49.1935,-24.5967)(0.,0.)
\ArrowLine(80.,-40.)(49.1935,-24.5967)
\Photon(56.,28.)(49.1935,0.){2.4}{2}
\Photon(24.,12.)(49.1935,0.){-2.4}{2}
\Photon(49.1935,0.)(49.1935,-24.5967){2.4}{2}
\Vertex(56.,28.){2}
\Vertex(24.,12.){2}
\Vertex(49.1935,0.){2}
\Vertex(49.1935,-24.5967){2}
\Text(84.,42.)[l]{$\leg{1}$}
\Text(84.,-42.)[l]{$\leg{2}$}
\Text(56.5825,13.3573)[l]{$\scriptstyle{a_1}$}
\Text(56.5825,-13.3573)[l]{$\scriptstyle{a_3}$}
\Text(35.307,5.78034)[tr]{$\scriptstyle{a_2}$}
\Text(41.5019,21.9679)[br]{}
\Text(32.,-16.)[tr]{}
\Text(19.515,10.9162)[br]{}
\end{picture}
}

\newcommand{\diagIV}{
\begin{picture}(120.,104.)(-28.,-52.)
\Gluon(-35.7771,0.)(0.,0.){3.2}{4}
\Vertex(0.,0.){2}
\ArrowLine(0.,0.)(44.7214,22.3607)
\ArrowLine(44.7214,22.3607)(80.,40.)
\ArrowLine(24.,-12.)(0.,0.)
\ArrowLine(56.,-28.)(24.,-12.)
\ArrowLine(80.,-40.)(56.,-28.)
\Photon(56.,-28.)(44.7214,0.){2.4}{2}
\Photon(24.,-12.)(44.7214,0.){2.4}{2}
\Photon(44.7214,0.)(44.7214,22.3607){2.4}{2}
\Vertex(56.,-28.){2}
\Vertex(24.,-12.){2}
\Vertex(44.7214,0.){2}
\Vertex(44.7214,22.3607){2}
\Text(84.,42.)[l]{$\leg{1}$}
\Text(84.,-42.)[l]{$\leg{2}$}
\Text(59.194,-13.9738)[l]{$\scriptstyle{a_1}$}
\Text(54.8415,12.9463)[l]{$\scriptstyle{a_3}$}
\Text(31.0949,-3.6205)[br]{$\scriptstyle{a_2}$}
\Text(41.5019,-21.9679)[br]{}
\Text(32.,16.)[tr]{}
\Text(19.515,-10.9162)[br]{}
\end{picture}
}

\newcommand{\diagV}{
\begin{picture}(120.,104.)(-28.,-52.)
\Gluon(-35.7771,0.)(0.,0.){3.2}{4}
\Vertex(0.,0.){2}
\ArrowLine(0.,0.)(16.,8.)
\ArrowLine(16.,8.)(44.,22.)
\ArrowLine(44.,22.)(64.,32.)
\ArrowLine(64.,32.)(80.,40.)
\ArrowLine(64.,-32.)(0.,0.)
\ArrowLine(80.,-40.)(64.,-32.)
\Photon(64.,32.)(64.,-32.){2.4}{3.5}
\PhotonArc(30.,15.)(15.6525,26.5651,206.565){2}{4}
\Vertex(64.,-32.){2}
\Vertex(16.,8.){2}
\Vertex(44.,22.){2}
\Vertex(64.,32.){2}
\Text(84.,42.)[l]{$\leg{1}$}
\Text(84.,-42.)[l]{$\leg{2}$}
\Text(71.5542,0.)[l]{$\scriptstyle{a_1}$}
\Text(23.8885,35.1332)[br]{$\scriptstyle{a_2}$}
\Text(41.5019,21.9679)[br]{}
\Text(42.,-21.)[tr]{}
\Text(19.515,10.9162)[br]{}
\Text(20.,-10.)[tr]{}
\end{picture}
}

\newcommand{\diagVI}{
\begin{picture}(120.,104.)(-28.,-52.)
\Gluon(-35.7771,0.)(0.,0.){3.2}{4}
\Vertex(0.,0.){2}
\ArrowLine(0.,0.)(64.,32.)
\ArrowLine(64.,32.)(80.,40.)
\ArrowLine(16.,-8.)(0.,0.)
\ArrowLine(44.,-22.)(16.,-8.)
\ArrowLine(64.,-32.)(44.,-22.)
\ArrowLine(80.,-40.)(64.,-32.)
\Photon(64.,32.)(64.,-32.){2.4}{3.5}
\PhotonArc(30.,-15.)(15.6525,-206.565,-26.5651){2}{4}
\Vertex(64.,-32.){2}
\Vertex(16.,-8.){2}
\Vertex(44.,-22.){2}
\Vertex(64.,32.){2}
\Text(84.,42.)[l]{$\leg{1}$}
\Text(84.,-42.)[l]{$\leg{2}$}
\Text(71.5542,0.)[l]{$\scriptstyle{a_1}$}
\Text(18.7867,-38.1059)[br]{$\scriptstyle{a_2}$}
\Text(41.5019,21.9679)[br]{}
\Text(42.,-21.)[tr]{}
\Text(19.515,10.9162)[br]{}
\Text(20.,-10.)[tr]{}
\end{picture}
}

\newcommand{\diagVII}{
\begin{picture}(120.,104.)(-28.,-52.)
\Gluon(-35.7771,0.)(0.,0.){3.2}{4}
\Vertex(0.,0.){2}
\ArrowLine(0.,0.)(24.,12.)
\ArrowLine(24.,12.)(40.,20.)
\ArrowLine(40.,20.)(56.,28.)
\ArrowLine(56.,28.)(80.,40.)
\ArrowLine(40.,-20.)(0.,0.)
\ArrowLine(80.,-40.)(40.,-20.)
\Photon(40.,20.)(40.,-20.){2.4}{3.5}
\PhotonArc(40.,20.)(17.8885,26.5651,206.565){2}{4}
\Vertex(40.,-20.){2}
\Vertex(24.,12.){2}
\Vertex(40.,20.){2}
\Vertex(56.,28.){2}
\Text(84.,42.)[l]{$\leg{1}$}
\Text(84.,-42.)[l]{$\leg{2}$}
\Text(49.1935,0.)[l]{$\scriptstyle{a_1}$}
\Text(32.1994,42.9325)[br]{$\scriptstyle{a_2}$}
\Text(41.5019,21.9679)[br]{}
\Text(42.,-21.)[tr]{}
\Text(19.515,10.9162)[br]{}
\Text(20.,-10.)[tr]{}
\end{picture}
}

\newcommand{\diagVIII}{
\begin{picture}(120.,104.)(-28.,-52.)
\Gluon(-35.7771,0.)(0.,0.){3.2}{4}
\Vertex(0.,0.){2}
\ArrowLine(0.,0.)(40.,20.)
\ArrowLine(40.,20.)(80.,40.)
\ArrowLine(24.,-12.)(0.,0.)
\ArrowLine(40.,-20.)(24.,-12.)
\ArrowLine(56.,-28.)(40.,-20.)
\ArrowLine(80.,-40.)(56.,-28.)
\Photon(40.,20.)(40.,-20.){2.4}{3.5}
\PhotonArc(40.,-20.)(17.8885,-206.565,-26.5651){2}{4}
\Vertex(40.,-20.){2}
\Vertex(24.,-12.){2}
\Vertex(56.,-28.){2}
\Vertex(40.,20.){2}
\Text(84.,42.)[l]{$\leg{1}$}
\Text(84.,-42.)[l]{$\leg{2}$}
\Text(49.1935,0.)[l]{$\scriptstyle{a_1}$}
\Text(30.1749,-44.3787)[br]{$\scriptstyle{a_2}$}
\Text(41.5019,21.9679)[br]{}
\Text(42.,-21.)[tr]{}
\Text(19.515,10.9162)[br]{}
\Text(20.,-10.)[tr]{}
\end{picture}
}

\newcommand{\diagIX}{
\begin{picture}(120.,104.)(-28.,-52.)
\Gluon(-35.7771,0.)(0.,0.){3.2}{4}
\Vertex(0.,0.){2}
\ArrowLine(0.,0.)(56.,28.)
\ArrowLine(56.,28.)(80.,40.)
\ArrowLine(56.,-28.)(0.,0.)
\ArrowLine(80.,-40.)(56.,-28.)
\Photon(56.,28.)(56.,10.5064){2.4}{2.5}
\Photon(56.,-10.5064)(56.,-28.){2.4}{2.5}
\ArrowArc(56.9771,0.)(11.7466,-90.,90.)
\ArrowArc(56.9771,0.)(11.7466,90.,270.)
\Vertex(56.,-28.){2}
\Vertex(56.,28.){2}
\Vertex(56.,10.5064){2}
\Vertex(56.,-10.5064){2}
\Text(84.,42.)[l]{$\leg{1}$}
\Text(84.,-42.)[l]{$\leg{2}$}
\Text(64.4276,21.6767)[lu]{$\scriptstyle{a_1}$}
\Text(63.3537,-24.6395)[lb]{$\scriptstyle{a_4}$}
\Text(74.2375,0.)[l]{$\scriptstyle{\Psi_i}$}
\Text(41.1437,0.)[r]{$\scriptstyle{\Psi_j}$}
\Text(41.5019,21.9679)[br]{}
\Text(42.,-21.)[tr]{}
\Text(19.515,10.9162)[br]{}
\Text(20.,-10.)[tr]{}
\end{picture}
}

\newcommand{\diagX}{
\begin{picture}(120.,104.)(-28.,-52.)
\Gluon(-35.7771,0.)(0.,0.){3.2}{4}
\Vertex(0.,0.){2}
\ArrowLine(0.,0.)(56.,28.)
\ArrowLine(56.,28.)(80.,40.)
\ArrowLine(56.,-28.)(0.,0.)
\ArrowLine(80.,-40.)(56.,-28.)
\Photon(56.,28.)(56.,10.5064){2.4}{2.5}
\Photon(56.,-10.5064)(56.,-28.){2.4}{2.5}
\PhotonArc(56.9771,0.)(11.7466,-101.459,258.541){-2}{8}
\Vertex(56.,-28.){2}
\Vertex(56.,28.){2}
\Vertex(56.,10.5064){2}
\Vertex(56.,-10.5064){2}
\Text(84.,42.)[l]{$\leg{1}$}
\Text(84.,-42.)[l]{$\leg{2}$}
\Text(64.4276,21.6767)[lu]{$\scriptstyle{a_1}$}
\Text(76.0263,0.)[l]{$\scriptstyle{a_2}$}
\Text(38.4604,0.)[r]{$\scriptstyle{a_3}$}
\Text(63.3537,-24.6395)[lb]{$\scriptstyle{a_4}$}
\Text(41.5019,21.9679)[br]{}
\Text(42.,-21.)[tr]{}
\Text(19.515,10.9162)[br]{}
\Text(20.,-10.)[tr]{}
\end{picture}
}

\newcommand{\diagXI}{
\begin{picture}(120.,104.)(-28.,-52.)
\Gluon(-35.7771,0.)(0.,0.){3.2}{4}
\Vertex(0.,0.){2}
\ArrowLine(0.,0.)(56.,28.)
\ArrowLine(56.,28.)(80.,40.)
\ArrowLine(56.,-28.)(0.,0.)
\ArrowLine(80.,-40.)(56.,-28.)
\Photon(56.,28.)(56.,10.5064){2.4}{2.5}
\Photon(56.,-10.5064)(56.,-28.){2.4}{2.5}
\DashArrowArc(56.9771,0.)(11.7466,90.,-90.){1}
\DashArrowArc(56.9771,0.)(11.7466,270.,90.){1}
\Vertex(56.,-28.){2}
\Vertex(56.,28.){2}
\Vertex(56.,10.5064){2}
\Vertex(56.,-10.5064){2}
\Text(84.,42.)[l]{$\leg{1}$}
\Text(84.,-42.)[l]{$\leg{2}$}
\Text(64.4276,21.6767)[lu]{$\scriptstyle{a_1}$}
\Text(73.343,0.)[l]{$\scriptstyle{u^{a_2}}$}
\Text(41.1437,0.)[r]{$\scriptstyle{u^{a_3}}$}
\Text(63.3537,-24.6395)[lb]{$\scriptstyle{a_4}$}
\Text(41.5019,21.9679)[br]{}
\Text(42.,-21.)[tr]{}
\Text(19.515,10.9162)[br]{}
\Text(20.,-10.)[tr]{}
\end{picture}
}

\newcommand{\diagXII}{
\begin{picture}(120.,104.)(-28.,-52.)
\Gluon(-35.7771,0.)(0.,0.){3.2}{4}
\Vertex(0.,0.){2}
\ArrowLine(0.,0.)(56.,28.)
\ArrowLine(56.,28.)(80.,40.)
\ArrowLine(56.,-28.)(0.,0.)
\ArrowLine(80.,-40.)(56.,-28.)
\Photon(56.,28.)(56.,10.5064){2.4}{2.5}
\Photon(56.,-10.5064)(56.,-28.){2.4}{2.5}
\DashCArc(56.9771,0.)(11.7466,-90.,90.){3}
\DashCArc(56.9771,0.)(11.7466,90.,270.){3}
\Vertex(56.,-28.){2}
\Vertex(56.,28.){2}
\Vertex(56.,10.5064){2}
\Vertex(56.,-10.5064){2}
\Text(84.,42.)[l]{$\leg{1}$}
\Text(84.,-42.)[l]{$\leg{2}$}
\Text(64.4276,21.6767)[lu]{$\scriptstyle{a_1}$}
\Text(71.5542,0.)[l]{$\scriptstyle{\Phi_{i_2}}$}
\Text(42.9325,0.)[r]{$\scriptstyle{\Phi_{i_3}}$}
\Text(63.3537,-24.6395)[lb]{$\scriptstyle{a_4}$}
\Text(41.5019,21.9679)[br]{}
\Text(42.,-21.)[tr]{}
\Text(19.515,10.9162)[br]{}
\Text(20.,-10.)[tr]{}
\end{picture}
}

\newcommand{\diagXV}{
\begin{picture}(120.,104.)(-28.,-52.)
\Gluon(-35.7771,0.)(0.,0.){3.2}{4}
\Vertex(0.,0.){2}
\ArrowLine(0.,0.)(56.,28.)
\ArrowLine(56.,28.)(80.,40.)
\ArrowLine(56.,-28.)(0.,0.)
\ArrowLine(80.,-40.)(56.,-28.)
\Photon(56.,28.)(56.,10.5064){-2.4}{2.5}
\Photon(56.,-10.5064)(56.,-28.){-2.4}{2.5}
\PhotonArc(56.9771,0.)(11.7466,90.,270.){-2}{3.5}
\DashCArc(56.9771,0.)(11.7466,-90.,90.){3}
\Vertex(56.,-28.){2}
\Vertex(56.,28.){2}
\Vertex(56.,10.5064){2}
\Vertex(56.,-10.5064){2}
\Text(84.,42.)[l]{$\leg{1}$}
\Text(84.,-42.)[l]{$\leg{2}$}
\Text(64.4276,21.6767)[lu]{$\scriptstyle{a_1}$}
\Text(76.0263,0.)[l]{$\scriptstyle{\Phi_{i_2}}$}
\Text(38.4604,0.)[r]{$\scriptstyle{a_3}$}
\Text(63.3537,-24.6395)[lb]{$\scriptstyle{a_4}$}
\Text(41.5019,21.9679)[br]{}
\Text(42.,-21.)[tr]{}
\Text(19.515,10.9162)[br]{}
\Text(20.,-10.)[tr]{}
\end{picture}
}

\newcommand{\diagXVI}{
\begin{picture}(120.,104.)(-28.,-52.)
\Gluon(-35.7771,0.)(0.,0.){3.2}{4}
\Vertex(0.,0.){2}
\ArrowLine(0.,0.)(48.,24.)
\ArrowLine(48.,24.)(80.,40.)
\ArrowLine(48.,-24.)(0.,0.)
\ArrowLine(80.,-40.)(48.,-24.)
\Photon(48.,24.)(58.1378,0.){-2.4}{2.5}
\Photon(48.,-24.)(58.1378,0.){2.4}{2.5}
\PhotonArc(67.082,0.)(8.94427,-177.135,182.865){-2}{7}
\Vertex(48.,-24.){2}
\Vertex(48.,24.){2}
\Vertex(58.1378,0.){2}
\Text(84.,42.)[l]{$\leg{1}$}
\Text(84.,-42.)[l]{$\leg{2}$}
\Text(55.9503,18.8245)[lu]{$\scriptstyle{a_1}$}
\Text(80.4984,0.)[l]{$\scriptstyle{a_2}$}
\Text(55.0177,-21.3975)[lb]{$\scriptstyle{a_3}$}
\Text(41.5019,21.9679)[br]{}
\Text(42.,-21.)[tr]{}
\Text(19.515,10.9162)[br]{}
\Text(20.,-10.)[tr]{}
\end{picture}
}

\newcommand{\diagXVII}{
\begin{picture}(120.,104.)(-28.,-52.)
\Gluon(-35.7771,0.)(0.,0.){3.2}{4}
\Vertex(0.,0.){2}
\ArrowLine(0.,0.)(48.,24.)
\ArrowLine(48.,24.)(80.,40.)
\ArrowLine(48.,-24.)(0.,0.)
\ArrowLine(80.,-40.)(48.,-24.)
\Photon(48.,24.)(58.1378,0.){-2.4}{2.5}
\Photon(48.,-24.)(58.1378,0.){2.4}{2.5}
\DashCArc(67.082,0.)(8.94427,-177.135,182.865){3}
\Vertex(48.,-24.){2}
\Vertex(48.,24.){2}
\Vertex(58.1378,0.){2}
\Text(84.,42.)[l]{$\leg{1}$}
\Text(84.,-42.)[l]{$\leg{2}$}
\Text(55.9503,18.8245)[lu]{$\scriptstyle{a_1}$}
\Text(80.4984,0.)[l]{$\scriptstyle{\Phi_{i_2}}$}
\Text(55.0177,-21.3975)[lb]{$\scriptstyle{a_3}$}
\Text(41.5019,21.9679)[br]{}
\Text(42.,-21.)[tr]{}
\Text(19.515,10.9162)[br]{}
\Text(20.,-10.)[tr]{}
\end{picture}
}

\newcommand{\diagAZ}{
\begin{picture}(120.,104.)(-28.,-52.)
\Gluon(-35.7771,0.)(0.,0.){3.2}{4}
\Vertex(0.,0.){2}
\ArrowLine(0.,0.)(56.,28.)
\ArrowLine(56.,28.)(80.,40.)
\ArrowLine(56.,-28.)(0.,0.)
\ArrowLine(80.,-40.)(56.,-28.)
\Photon(56.,28.)(56.,10.5064){-2.4}{2.5}
\Photon(56.,-10.5064)(56.,-28.){-2.4}{2.5}
\GCirc(56.,0.){11}{0.8}
\Vertex(56.,-28.){2}
\Vertex(56.,28.){2}
\Text(84.,42.)[l]{$\leg{1}$}
\Text(84.,-42.)[l]{$\leg{2}$}
\Text(64.4276,21.6767)[lu]{$\scriptstyle{A}$}
\Text(76.0263,0.)[l]{}
\Text(38.4604,0.)[r]{}
\Text(63.9079,-23.1643)[lb]{$\scriptstyle{Z}$}
\Text(41.5019,21.9679)[br]{}
\Text(42.,-21.)[tr]{}
\Text(19.515,10.9162)[br]{}
\Text(20.,-10.)[tr]{}
\end{picture}
}


\newcommand{\Newcommand}{\newcommand}
\newcommand{\Nextline}{\nl}
\newcommand{\Frac}{\frac}
\newcommand{\Epsinv}[1]{\veps^{- #1}}
\newcommand{\Eps}[1]{\veps^{#1}}
\newcommand{\Right}{\right}
\newcommand{\EG}{\gamma_{\mathrm{E}}}

\Newcommand{\ResIunivP}{
-2\Epsinv{2}
-3\Epsinv{1}
}

\Newcommand{\ResIunivW}{
-L^2
-\frac{2}{3}L^3\Eps{}
-\frac{1}{4}L^4\Eps{2}
+3L
+\frac{3}{2}L^2\Eps{}
+\frac{1}{2}L^3\Eps{2}
}

\Newcommand{\ResIunivZ}{
-L^2
-\frac{2}{3}L^3\Eps{}
-\frac{1}{4}L^4\Eps{2}
+\left(3 + 2\,\LMZW\Right)L
+\left(\frac{3}{2} + 2\,\LMZW\Right)L^2\Eps{}
+\left(\frac{1}{2} + \LMZW\Right)L^3\Eps{2}
}

\Newcommand{\ResdeIunivZ}{
2L
+2L^2\Eps{}
+L^3\Eps{2}
}

\Newcommand{\ResIunivPP}{
4\Epsinv{4}
+12\Epsinv{3}
}

\Newcommand{\ResIunivWW}{
L^4
-6L^3
}

\Newcommand{\ResIunivPW}{
2L^2\Epsinv{2}
+\frac{4}{3}L^3\Epsinv{1}
+\frac{1}{2}L^4
-6L\Epsinv{2}
+L^3
}

\Newcommand{\ResdeIunivPZ}{
-4L\Epsinv{2}
-4L^2\Epsinv{1}
-2L^3
}

\Newcommand{\ResdeIunivWZ}{
-2L^3
}

\Newcommand{\ResNLLtermP}{
\frac{3}{2}\Epsinv{3}
\left(-2\,\LmuM\Right)\Epsinv{2}
\left({\LmuM}^2\Right)\Epsinv{1}
\left(\frac{-{\LmuM}^3}{3}\Right)
}

\Newcommand{\ResNLLtermW}{
-\frac{2}{3}L^3
\left(-\LmuM\Right)L^2
}

\Newcommand{\ResoneM}{
-L^2
-\frac{2}{3}L^3\Eps{}
-\frac{1}{4}L^4\Eps{2}
+\Epsinv{1}
+4L
+2L^2\Eps{}
+\frac{2}{3}L^3\Eps{2}
}

\Newcommand{\ResdeoneM}{
2L
+2L^2\Eps{}
+L^3\Eps{2}
}

\Newcommand{\ResdeoneZ}{
-2\Epsinv{2}
+L^2
+\frac{2}{3}L^3\Eps{}
+\frac{1}{4}L^4\Eps{2}
-4\Epsinv{1}
-4L
-2L^2\Eps{}
-\frac{2}{3}L^3\Eps{2}
}

\Newcommand{\ResIMM}{
\frac{1}{6}L^4
-L^2\Epsinv{1}
-2L^3
}

\Newcommand{\ResdeIMM}{
-\frac{2}{3}L^3
}

\Newcommand{\ResdeIMZ}{
-\frac{2}{3}L^3
}

\Newcommand{\ResdeIZM}{
2L^2\Epsinv{2}
+\frac{8}{3}L^3\Epsinv{1}
+\frac{11}{6}L^4
-2\Epsinv{3}
-\left(8 + 4\,\LrMII\Right)L\Epsinv{2}
\nl&&{}
-\left(7 + 8\,\LrMII\Right)L^2\Epsinv{1}
-\left(\frac{10}{3} + 8\,\LrMII\Right)L^3
}

\Newcommand{\ResdeIZZ}{
\Epsinv{4}
-\frac{1}{6}L^4
+2\Epsinv{3}
+L^2\Epsinv{1}
+2L^3
}

\Newcommand{\ResIIMM}{
\frac{1}{3}L^4
-\frac{8}{3}L^3
}

\Newcommand{\ResdeIIMM}{
-\frac{2}{3}L^3
}

\Newcommand{\ResdeIIMZ}{
-\frac{2}{3}L^3\Epsinv{1}
-\frac{7}{6}L^4
+\left(4 + 2\,\LrMI\Right)L^2\Epsinv{1}
+\left(\frac{20}{3} + \frac{10\,\LrMI}{3}\Right)L^3
}

\Newcommand{\ResdeIIZM}{
-\frac{2}{3}L^3\Epsinv{1}
-\frac{7}{6}L^4
+\left(4 + 2\,\LrMII\Right)L^2\Epsinv{1}
+\left(\frac{20}{3} + \frac{10\,\LrMII}{3}\Right)L^3
}

\Newcommand{\ResdeIIZZ}{
\Epsinv{4}
-\frac{1}{3}L^4
+4\Epsinv{3}
+\frac{8}{3}L^3
}

\Newcommand{\ResIIIMMM}{
\frac{1}{6}L^4
-3L^2\Epsinv{1}
-5L^3
}

\Newcommand{\ResdeIIIMMM}{
-\frac{1}{3}L^3
}

\Newcommand{\ResdeIIIZMM}{
-\frac{1}{3}L^3\Epsinv{1}
-\frac{7}{12}L^4
+\left(2 + \LrMIII\Right)L^2\Epsinv{1}
+\left(\frac{10}{3} + \frac{5\,\LrMIII}{3}\Right)L^3
}

\Newcommand{\ResdeIIIMZM}{
-\frac{1}{3}L^3
}

\Newcommand{\ResdeIIIMMZ}{
-\frac{1}{3}L^3\Epsinv{1}
-\frac{7}{12}L^4
-6\Epsinv{3}
-6L\Epsinv{2}
+\left(1 + \LrMI\Right)L^2\Epsinv{1}
\nl&&{}
+\left(\frac{17}{3} + \frac{5\,\LrMI}{3}\Right)L^3
}

\Newcommand{\ResIVMMM}{
\frac{1}{6}L^4
-3L^2\Epsinv{1}
-5L^3
}

\Newcommand{\ResdeIVMMM}{
-\frac{1}{3}L^3
}

\Newcommand{\ResdeIVZMM}{
-\frac{1}{3}L^3\Epsinv{1}
-\frac{7}{12}L^4
+\left(2 + \LrMIII\Right)L^2\Epsinv{1}
+\left(\frac{10}{3} + \frac{5\,\LrMIII}{3}\Right)L^3
}

\Newcommand{\ResdeIVMZM}{
-\frac{1}{3}L^3
}

\Newcommand{\ResdeIVMMZ}{
-\frac{1}{3}L^3\Epsinv{1}
-\frac{7}{12}L^4
-6\Epsinv{3}
-6L\Epsinv{2}
+\left(1 + \LrMI\Right)L^2\Epsinv{1}
+\left(\frac{17}{3} + \frac{5\,\LrMI}{3}\Right)L^3
}

\Newcommand{\ResVMM}{
L^2\Epsinv{1}
+L^3
}

\Newcommand{\ResdeVMM}{0
}

\Newcommand{\ResdeVMZ}{0
}

\Newcommand{\ResdeVZM}{
2\Epsinv{3}
+2L\Epsinv{2}
+L^2\Epsinv{1}
+\frac{1}{3}L^3
}

\Newcommand{\ResdeVZZ}{
\Epsinv{3}
-L^2\Epsinv{1}
-L^3
}

\Newcommand{\ResVIMM}{
L^2\Epsinv{1}
+L^3
}

\Newcommand{\ResdeVIMM}{0
}

\Newcommand{\ResdeVIMZ}{0
}

\Newcommand{\ResdeVIZM}{
2\Epsinv{3}
+2L\Epsinv{2}
+L^2\Epsinv{1}
+\frac{1}{3}L^3
}

\Newcommand{\ResdeVIZZ}{
\Epsinv{3}
-L^2\Epsinv{1}
-L^3
}

\Newcommand{\ResVIIMM}{
-L^2\Epsinv{1}
-L^3
}

\Newcommand{\ResdeVIIMM}{0
}

\Newcommand{\ResdeVIIMZ}{0
}

\Newcommand{\ResdeVIIZM}{
-2\Epsinv{3}
-2L\Epsinv{2}
-L^2\Epsinv{1}
-\frac{1}{3}L^3
}

\Newcommand{\ResdeVIIZZ}{
-\Epsinv{3}
+L^2\Epsinv{1}
+L^3
}

\Newcommand{\ResVIIIMM}{
-L^2\Epsinv{1}
-L^3
}

\Newcommand{\ResdeVIIIMM}{0
}

\Newcommand{\ResdeVIIIMZ}{0
}

\Newcommand{\ResdeVIIIZM}{
-2\Epsinv{3}
-2L\Epsinv{2}
-L^2\Epsinv{1}
-\frac{1}{3}L^3
}

\Newcommand{\ResdeVIIIZZ}{
-\Epsinv{3}
+L^2\Epsinv{1}
+L^3
}

\Newcommand{\ResdeIXMZZM}{0
}

\Newcommand{\ResdeIXMZZZ}{0
}

\Newcommand{\ResdeIXZZZM}{0
}

\Newcommand{\ResdeIXZZZZ}{
\frac{2}{3}\Epsinv{3}
-\frac{4}{3}L^2\Epsinv{1}
-\frac{16}{9}L^3
}

\Newcommand{\ResIXMMMM}{
\frac{4}{3}L^2\Epsinv{1}
+\frac{16}{9}L^3
}

\Newcommand{\ResdeIXMMMM}{0
}

\Newcommand{\ResdeIXMMMZ}{
8\Epsinv{3}
+8L\Epsinv{2}
-\frac{16}{3}L^3
}

\Newcommand{\ResdeIXZMMM}{
8\Epsinv{3}
+8L\Epsinv{2}
-\frac{16}{3}L^3
}

\Newcommand{\ResdeIXZMMZ}{
\frac{8}{3}\Epsinv{3}
+\frac{8}{3}L\Epsinv{2}
-\frac{16}{9}L^3
}

\Newcommand{\ResdeIXMZMM}{0
}

\Newcommand{\ResdeIXMMZM}{0
}

\Newcommand{\ResIXmMMMM}{0
}

\Newcommand{\ResdeIXmMZZM}{0
}

\Newcommand{\ResdeIXmMMMM}{0
}

\Newcommand{\ResdeIXmMMMZ}{
-8\Epsinv{3}
-8L\Epsinv{2}
+\frac{16}{3}L^3
}

\Newcommand{\ResdeIXmZMMM}{
-8\Epsinv{3}
-8L\Epsinv{2}
+\frac{16}{3}L^3
}

\Newcommand{\ResdeIXmZMMZ}{0
}

\Newcommand{\ResXMMMM}{
\frac{10}{3}L^2\Epsinv{1}
+\frac{40}{9}L^3
}

\Newcommand{\ResdeXMMMM}{0
}

\Newcommand{\ResdeXZMMM}{
-16\Epsinv{3}
-16L\Epsinv{2}
+\frac{32}{3}L^3
}

\Newcommand{\ResdeXMZMM}{0
}

\Newcommand{\ResdeXMMZM}{0
}

\Newcommand{\ResdeXMMMZ}{
-16\Epsinv{3}
-16L\Epsinv{2}
+\frac{32}{3}L^3
}

\Newcommand{\ResdeXZMMZ}{
\frac{20}{3}\Epsinv{3}
+\frac{20}{3}L\Epsinv{2}
-\frac{40}{9}L^3
}

\Newcommand{\ResXIMMMM}{
\frac{1}{12}L^2\Epsinv{1}
+\frac{1}{9}L^3
}

\Newcommand{\ResdeXIMMMM}{
-\frac{13}{4}L^2\Epsinv{1}
-\frac{13}{3}L^3
}

\Newcommand{\ResdeXIZMMM}{
\Epsinv{3}
+L\Epsinv{2}
-\frac{2}{3}L^3
}

\Newcommand{\ResdeXIMZMM}{0
}

\Newcommand{\ResdeXIMMZM}{0
}

\Newcommand{\ResdeXIMMMZ}{
\Epsinv{3}
+L\Epsinv{2}
-\frac{2}{3}L^3
}

\Newcommand{\ResdeXIZMMZ}{
\frac{1}{6}\Epsinv{3}
+\frac{1}{6}L\Epsinv{2}
-\frac{1}{9}L^3
}

\Newcommand{\ResXIIMZZM}{
\frac{1}{3}L^2\Epsinv{1}
+\frac{4}{9}L^3
}

\Newcommand{\ResdeXIIMZZM}{0
}

\Newcommand{\ResdeXIIMZZZ}{0
}

\Newcommand{\ResdeXIIZZZM}{0
}

\Newcommand{\ResdeXIIZZZZ}{
\frac{1}{6}\Epsinv{3}
-\frac{1}{3}L^2\Epsinv{1}
-\frac{4}{9}L^3
}

\Newcommand{\ResXIIMMMM}{
\frac{1}{3}L^2\Epsinv{1}
+\frac{4}{9}L^3
}

\Newcommand{\ResdeXIIMMMM}{0
}

\Newcommand{\ResdeXIIMMMZ}{
4\Epsinv{3}
+4L\Epsinv{2}
-\frac{8}{3}L^3
}

\Newcommand{\ResdeXIIZMMM}{
4\Epsinv{3}
+4L\Epsinv{2}
-\frac{8}{3}L^3
}

\Newcommand{\ResdeXIIZMMZ}{
\frac{2}{3}\Epsinv{3}
+\frac{2}{3}L\Epsinv{2}
-\frac{4}{9}L^3
}

\Newcommand{\ResXVMMMM}{0
}

\Newcommand{\ResdeXVMMMM}{0
}

\Newcommand{\ResdeXVZMMM}{
-2\Epsinv{3}
-2L\Epsinv{2}
+\frac{4}{3}L^3
}

\Newcommand{\ResdeXVMMZM}{0
}

\Newcommand{\ResdeXVMMMZ}{
-2\Epsinv{3}
-2L\Epsinv{2}
+\frac{4}{3}L^3
}

\Newcommand{\ResdeXVZMMZ}{0
}

\Newcommand{\ResXVIMMM}{0
}

\Newcommand{\ResdeXVIMMM}{0
}

\Newcommand{\ResdeXVIZMM}{
-12\Epsinv{3}
-12L\Epsinv{2}
+8L^3
}

\Newcommand{\ResdeXVIMZM}{0
}

\Newcommand{\ResdeXVIMMZ}{
-12\Epsinv{3}
-12L\Epsinv{2}
+8L^3
}

\Newcommand{\ResdeXVIZMZ}{0
}

\Newcommand{\ResXVIIMMM}{0
}

\Newcommand{\ResdeXVIIMMM}{0
}

\Newcommand{\ResdeXVIIZMM}{
-2\Epsinv{3}
-2L\Epsinv{2}
+\frac{4}{3}L^3
}

\Newcommand{\ResdeXVIIMMZ}{
-2\Epsinv{3}
-2L\Epsinv{2}
+\frac{4}{3}L^3
}

\Newcommand{\ResdeXVIIZMZ}{0
}


\thispagestyle{empty} 


\thispagestyle{empty}
\def\thefootnote{\fnsymbol{footnote}}
\setcounter{footnote}{1}
\null
\draftdate\hfill   TTP04-01\\
\strut\hfill SFB/CPP-04-43 \\
\strut\hfill hep-ph/0401087
\vskip 0cm
\vfill
\begin{center}
{\Large \bf
Next-to-leading mass singularities
in two-loop electroweak
singlet form factors
\par}
 \vskip 1em
{\large
{\sc S.~Pozzorini\footnote{pozzorin@particle.uni-karlsruhe.de} }}
\\[.5cm]
{\it Institut f\"ur Theoretische Teilchenphysik, 
Universit\"at Karlsruhe \\
D-76128 Karlsruhe, Germany}
\par
\end{center}\par
\vskip 1.0cm 
\vfill 
{\bf Abstract:} \par

We consider 
virtual electroweak corrections
to the form factors for  
massless chiral fermions coupling to an $\SUtwo\times \Uone$ singlet 
gauge boson 
in the asymptotic region $s\gg\MW^2\sim\MZ^2$, 
where the invariant mass $s$ of the external gauge boson 
is much higher than the weak-boson mass scale.
Using the sector-decomposition method
we compute 
mass singularities, which  arise as logarithms 
of $s/\MW^2$ and  $1/\veps$ poles in $D=4-2\veps$ dimensions,
to one- and two-loop 
next-to-leading-logarithmic accuracy.
In this approximation we include all contributions
of order
$\alpha^l\veps^{k}\log^{j+k}(s/\MW^2)$, with $l=1,2$
and $j=2l,2l-1$.
We find that the electroweak two-loop 
leading- and next-to-leading-logarithmic
mass singularities can be written 
in a form that corresponds to a generalization of 
Catani's formula for massless QCD.

\par
\vskip 1cm
\noindent
January 2004 
\par
\null
\setcounter{page}{0}
\clearpage
\def\thefootnote{\arabic{footnote}}
\setcounter{footnote}{0}

\newcommand{\univfact}[2]{I(#1,#2)}
\newcommand{\NLLfact}[3]{J(#1,#2,#3)}
\newcommand{\FF}[2]{F_{#1}^{(#2)}}
\newcommand{\DD}[1]{D_{#1}}
\newcommand{\deDD}[1]{\Delta D_{#1}}
\newcommand{\gb}{g_1}
\newcommand{\gw}{g_2}
\newcommand{\CA}{C_A}
\newcommand{\Rep}{R}
\newcommand{\groupa}{\Uone}
\newcommand{\groupb}{\SUtwo}
\newcommand{\wbos}{V}
\newcommand{\coeffsymbol}{K}

\newcommand{\one}{1}
\newcommand{\two}{2}
\newcommand{\thr}{3}
\newcommand{\fiv}{4}
\newcommand{\sev}{5}
\newcommand{\ten}{6}
\newcommand{\sixteen}{7}
\newcommand{\fif}{8}
\newcommand{\twe}{9}
\newcommand{\seventeen}{10}
\newcommand{\nin}{11}

\newcommand{\projector}{{\mathcal{P}}}
\newcommand{\normfact}{N_{\veps}}
\newcommand{\diag}[1]{\mathrm{Diag}_{#1}}
\newcommand{\mrat}{r}
\newcommand{\lr}[1]{\mathrm{l}_{#1}}
\newcommand{\LMZW}{\lr{\mathrm{Z}}}
\newcommand{\LmuM}{\lr{\mu_\mathrm{w}}}
\newcommand{\Lmue}{\lr{\mu_e}}
\newcommand{\Lmui}{\lr{\mu_i}}
\newcommand{\LrMI}{\lr{1}}
\newcommand{\LrMII}{\lr{2}}
\newcommand{\LrMIII}{\lr{3}}
\newcommand{\LrMIV}{\lr{4}}
\newcommand{\equaldiag}{\hspace{-3mm}=}
\newcommand{\dimtwo}{d_2}

\newcommand{\pmass}{0}

\section{Introduction}
\label{se:intro}
Future colliders, such as the LHC
\cite{Haywood:1999qg} or an $\Pep\Pem$ Linear Collider (LC)
\cite{Aguilar-Saavedra:2001rg,Abe:2001wn,Abe:2001gc},
will investigate 
interactions between the constituents of the Standard Model 
in a new window of energies 
above the present frontier of 
few hundred GeV
up to the TeV scale and with very high luminosities.
On the one hand, this window will 
provide new insights into the mechanism of electroweak (EW) 
symmetry-breaking 
through either the discovery of 
the Higgs boson and the measurement of its couplings 
or the observation of strongly interacting vector bosons.
On the other hand, it will open new perspectives 
for precision tests of the Standard Model.

From the point of view of 
theoretical predictions,
since the size of radiative corrections 
grows with energy,
multi-loop calculations will play an increasingly important role,
especially  in order to achieve the permille-level accuracy 
required by a machine such as a LC.
In this context, the behaviour of the EW corrections is particularly
interesting.
Indeed, if one considers  the range of energies much higher than the weak-boson 
mass scale, $\sqrt{s}\gg\MW$,
and performs an expansion in
$\MW/\sqrt{s}$,
the leading terms of this expansion
arise as a tower of logarithms
\cite{Kuroda:1991wn}, 
\beq\label{logform}
\alpha^l\log^{j}{\left(\frac{s}{\MW^2}\right)},
\qquad\mbox{with}\quad
\quad 0\le j \le 2l,
\eeq
which  clearly enhance the EW corrections.
At $\sqrt{s}=1 \TeV$, these logarithms
yield one-loop EW corrections of tens of percent
with a considerable impact on LHC observables,
and two-loop corrections of a few percent,
which must be under control at a LC.

At $l$-loop level,
the leading logarithms (LL's), also known as Sudakov
logarithms \cite{Sudakov:1954sw}, have power
$j=2l$, and  the subleading terms with $j=2l-1,2l-2,\dots$ are denoted
as next-to-leading logarithms (NLL's),
next-to-next-to-leading logarithms (NNLL's), and so on.
Analogously to mass singularities observed in QED and QCD,
the EW logarithms  arise from soft/collinear emission
of virtual or real particles off initial or final-state particles.
However, in contrast to QED and QCD,
the EW logarithms do not cancel in physical 
observables
since the weak-boson masses provide a physical cut-off
and there is no need to
include real Z- and W-boson bremsstrahlung.
Moreover, 
even observables that are 
fully inclusive with respect to  Z- and W-bremsstrahlung
are affected by EW  LL's
that  violate the Bloch--Nordsieck theorem \cite{Ciafaloni:2000df},
owing to the fact that 
the scattering particles (fermions or gauge bosons)  
carry non-abelian weak-isospin 
charges.

At one loop, the EW LL's and NLL's 
are now well understood.
On the one hand, 
explicit diagrammatic calculations for many $2\to2$
processes exist
\cite{Beenakker:1993tt,Ciafaloni:1999xg,Beccaria:2000fk,Layssac:2001ur}.
On the other hand, it has been 
proven that these logarithms are universal,
and general results have been given for arbitrary
processes that are not mass-suppressed at high energies
\cite{Denner:2001jv,Pozzorini:rs}. 
For  gauge-boson pair production at the LHC
it was shown that,  in the 
transverse-momentum region above few-hundred GeV, the EW logarithmic  corrections can become of the order
of the QCD corrections \cite{Accomando:2001fn}.

At two loops, the situation is more difficult, since 
the analytical evaluation of 
Feynman diagrams with many external legs 
and different internal masses is a highly non-trivial task.
At present, there are  good prospects 
to compute  vertex diagrams 
that depend on one single dimensionless parameter
$\MW^2/s$ \cite{Aglietti:2003yc}. However, 
exact analytical results for general
$2\to2$ processes seem to be out of reach.
It is therefore reasonable either to 
solve the problem numerically, 
as for instance in  \citere{Passarino:2001wv},
or to use analytical approximations.

Two different strategies have been adopted 
in order to treat  the higher-order EW logarithms 
analytically. 
On the one hand, 
evolution equations,
which are well known in 
QED and QCD, have been applied to the 
EW theory in order to resum
the one-loop logarithms\footnote{
In order to resum the LL's and NLL's 
it is sufficient to
determine the kernel of the evolution equations
to one-loop accuracy.
However, starting from the 
NNLL's also the two-loop contributions to the 
$\beta$-function and to the anomalous dimensions 
are needed.
}.
On the other hand, explicit diagrammatic 
calculations have been performed
using, for instance, the Sudakov method or the eikonal approximation.

Fadin \etal\ \cite{Fadin:2000bq} have
resummed the EW leading-logarithmic (LL) corrections to arbitrary
matrix elements by means of the  infrared evolution equation (IREE).
K\"uhn \etal\ have resummed the EW logarithmic
corrections to massless 4-fermion processes
up to the NNLL's \cite{Kuhn:2000nn}.
At the TeV scale they found that,
for these  particular processes,
the leading and subleading logarithms have similar size 
and alternating signs, which gives rise  
to large cancellations.
This means that, depending on the process, 
it might be necessary to compute the complete tower of logarithms
\refeq{logform}.
A  prescription for the resummation of the 
next-to-leading-logarithmic (NLL) corrections to arbitrary processes 
has been proposed by Melles \cite{Melles:2001gw,Melles:2001dh}.

All the above  resummations are based on the 
idea of splitting the EW theory
into two regimes, both  with exact gauge symmetry.
This idea was first formulated in the context of the 
IREE \cite{Fadin:2000bq}, which  describes the all-order LL dependence of
matrix elements with respect to the transverse-momentum cut-off
$\mu_\perp$ within symmetric gauge theories.
In practice, for  the regime $\sqrt{s}\ge\mu_\perp\ge \MW$ 
all EW gauge bosons are supposed to behave as if they would have the same
mass $\MW$ and one assumes $\SUtwo\times\Uone$ symmetry,
whereas in the regime $\MW\ge \mu_\perp$
the weak bosons are supposed to be frozen out
and $\Uone_\elm$ symmetry is assumed.
This latter regime describes the effects originating from the mass gap
in the gauge sector, \ie from the fact that the 
(massless) photon is much lighter 
than the weak bosons.

We note that in the above resummations,
apart from the splitting of the evolution into two regimes,
no other  effects from
spontaneous symmetry breaking are considered.
In particular, the following assumptions
are explicitly or implicitly made.
\begin{itemize}
\item[(i)] 
In the 
massless limit $\MW^2/s\to 0$,
all couplings with mass dimension,
 which originate from symmetry breaking, are neglected.

\item[(ii)] 
The weak-boson masses are 
introduced in the corresponding 
propagators
as regulators of soft
and collinear  singularities from  W and Z bosons
without spontaneous symmetry breaking. 
Since these masses are of the same order, 
one considers $\MW=\MZ$.

\item[(iii)] The regimes above and below the EW scale 
are treated as an
unmixed $\SUtwo\times\Uone$ theory 
and QED, respectively, and 
mixing effects in the gauge sector are neglected.

\end{itemize}
It is important 
to understand to which extend the above assumptions 
are legitimate 
and whether the resulting resummation
prescriptions are correct. This can be done 
by explicit diagrammatic two-loop
calculations based on the EW Lagrangian, where all 
effects related to spontaneous symmetry breaking are consistently 
taken into
account.  

At the LL level, these checks have been already completed. 
A calculation  of the massless fermionic
singlet form factor \cite{Melles:2000ed,Hori:2000tm} 
and then a Coulomb-gauge calculation for arbitrary processes 
\cite{Beenakker:2000kb}
have demonstrated that the EW LL's exponentiate 
as predicted by the IREE.
Also the angular-dependent subset of the NLL's
has been shown to exponentiate for arbitrary processes 
\cite{Denner:2003wi}
as anticipated in \citeres{Kuhn:2000nn,Melles:2001dh}.
The complete tower of two-loop logarithms
has been computed in \citere{Feucht:2003yx}
for the fermionic and scalar subsets of the 
corrections 
to a massless fermionic form factor
within an  abelian massive theory.

In this paper we present 
a first diagrammatic calculation where the complete set 
of EW LL and NLL corrections is taken into account.
We consider the one- and two-loop virtual EW corrections
to the form factors for  
massless chiral fermions coupling to an $\SUtwo\times \Uone$ singlet 
gauge boson.
All relevant loop integrals 
within the  't~Hooft--Feynman  gauge
are evaluated  in the asymptotic region $s\gg\MW^2\sim\MZ^2$ 
to NLL accuracy
 using the sector-decomposition method. 
In addition to the logarithms of the type \refeq{logform},
which arise from massive virtual particles,
we include also mass singularities 
from massless photons.
These latter
are regulated dimensionally and arise
as  $1/\veps$ poles in $D=4-2\veps$ dimensions.
In NLL approximation all mass singularities of the order
$\alpha^l\veps^{k}\log^{j+k}(s/\MW^2)$, with $j=2l,2l-1$, and  
$-j\le k\le 4-2l$, are taken into account.
We do not assume $\MZ=\MW$ and we include contributions depending
on $\log(\MZ/\MW)$.

The paper is organized as follows. 
In \refse{se:definitions}
we define our conventions
for the chiral form factors
and for the computation of 
the logarithmic EW corrections.
The one-loop results are presented in 
\refse{se:oneloop}.
In \refse{se:twoloop} we provide
the two-loop contributions from individual Feynman diagrams and
counterterms
and present the final result.
A detailed discussion is given in 
\refse{se:discussion}.
The definitions of the relevant one- and two-loop integrals 
as well as group-theoretical quantities related to the $\beta$-function
can be found in the appendices.

\newpage
\section{Definitions and conventions}
\label{se:definitions}
In this section we introduce 
chiral form factors and corresponding
projectors for  the vertex  involving 
an external $\SU(2)\times\Uone$ singlet gauge boson 
and a massless fermion--antifermion pair.
We define 
the conventions used to calculate the EW 
corrections,
the NLL approximation,
and a subtraction to isolate mass-gap effects.

\subsection{Chiral form factors in D dimensions}
Let us consider the vertex function
\beqar\label{diagramgeneric}
\vcenter{\hbox{
\diaggeneric
}}
\hspace{-13mm}&=&
\ri \bar{u}(p_1)F^\nu \, v(p_2)
\eeqar
with a fermion--antifermion pair
coupled to an external gauge boson.
This latter,
which might be for instance a gluon, 
is treated as an external field
that does not enter the virtual corrections.
Moreover,
it is assumed to be an $\SUtwo\times\Uone$ singlet,
which does not interact with the EW gauge bosons 
but only with fermions.

The outgoing fermionic momenta  and the invariant mass of the gauge boson
are denoted as 
$p_1,p_2$, and $s=(p_1+p_2)^2$, respectively.
We assume that the fermions are on-shell and massless.
In this case,
the above vertex can be parametrized
in terms of the left- and right-handed form factors
$F_-$ and $F_+$ as 
\beqar\label{formfactordecomp}
F^\nu=\gamma^\nu 
\sum_{\si=\pm}
\omega_\si F_\si,
\qquad\mbox{with}\qquad
\omega_\pm=\frac{1}{2}(1\pm\gamma^5),
\eeqar 
where indices corresponding to the weak isospin of the 
external fermions are implicitly understood.
The couplings of the singlet gauge boson
to the fermions are given by the tree-level form factors
and can be in general parity violating.

Since we compute the loop corrections within dimensional regularization,
we have to specify a prescription to handle 
$\gamma^5$ in $D=4-2\veps$ dimensions without 
violating the Ward identities of the external current.
In principle, a chiral anomaly could originate from
fermionic triangle subdiagrams. 
However, 
such triangle subdiagrams, and
the corresponding chiral anomaly, 
do not contribute to
the leading and next-to-leading mass singularities 
(see \refse{def:expasnions})
that we calculate
in this paper.
Therefore, we can safely adopt 
a dimensional regularization scheme with  
\beq\label{gammafive}
\{\gamma^\mu,\gamma^5\}=0.
\eeq
We note that this choice of scheme is not unique.
In principle, other prescriptions could be used which differ 
with respect to  \refeq{gammafive} by terms of order $\veps$.
As a consequence, the residues of the subleading $1/\veps$ poles 
in the infrared-divergent virtual and real corrections are scheme
dependent.
However, this ambiguity disappears 
in their infrared-finite combination, 
provided that the same prescription \refeq{gammafive}
is used for virtual and real corrections.

\subsection{Chiral projectors}
\label{se:projectors}
The left- and right-handed form factors  $F_\pm$ 
can be obtained by means of the following projectors 
\beqar\label{projectordef}
F_\si=
\projector_\si
\left(\hspace{8mm}
\vcenter{\hbox{
\diaggeneric
}}
\hspace{-13mm}\right)
&:=&
\frac{1}{(2-D)s} \Tr\left(\gamma_\nu \ps_1 F^\nu \omega_\si \ps_2\right)
.
\eeqar
When these projections $\projector_\si$ 
are applied to loop diagrams,  
the chiral projectors $\omega_\si$ 
can be easily combined 
with those originating from the 
couplings of the massless fermions to
virtual EW gauge bosons
$V=A,Z,W^\pm$,  
\beqar\label{chiralcouplingsdef}
\vcenter{\hbox{
\begin{picture}(110,100)(-50,-50)
\Text(-45,5)[lb]{$V_{\mu}$}
\Text(35,30)[cb]{$\bar{\Psi}_{i}$}
\Text(35,-30)[ct]{$\Psi_{j}$}
\Vertex(0,0){2}
\ArrowLine(0,0)(35,25)
\ArrowLine(35,-25)(0,0)
\Photon(0,0)(-45,0){2}{3}
\end{picture}}}
&=&\ri e \gamma_\mu \sum_{\rho=\pm}
\omega_\rho 
\left(I^{V}_{\rho}\right)_{\Psi_i\Psi_j}
.
\eeqar
Here $\Psi_{i}=u,d$ are the components of the fermionic doublet $\Psi$, 
and the generators $I^V_\rho$ 
[see \refeq{genmixing}]
are  isospin matrices in the representation corresponding 
to the chirality $\rho$,
\ie the  fundamental or singlet representation for 
left-handed ($\rho=-$) or right-handed ($\rho=+$) chirality,
respectively.

In practice, as a result of
chirality conservation for massless fermions, 
in each vertex \refeq{chiralcouplingsdef}
that is connected to the external gauge boson by means of a massless
fermionic line
one can replace $\sum_\rho \omega_\rho I^V_\rho$
by $I^V_\si$, \ie
the right- and left-handed form factors  \refeq{projectordef} 
receive contributions only from 
couplings with corresponding chirality $\rho=\si$.
After this simplification,
owing to the fact that 
the form factors depend only on two linearly independent external
momenta,
which cannot give rise to a totally antisymmetric tensor with four
indices,
the remaining $\gamma^5$ contribution from 
$\omega_\si$ drops out 
and one can set  $\omega_\si=1/2$ in 
\refeq{projectordef}.

\subsection{Perturbative and asymptotic expansions}
\label{def:expasnions}
As a convention, we write 
the perturbative expansion of the chiral form factors as 
\beqar\label{pertserie1}
F_\si&=&
\FF{\si}{0}
\left[
1+\sum_{l=1}^\infty
\left(\frac{\alpha}{4\pi}\right)^l
\normfact^l
\de \FF{\si}{l}
\right]
\quad\mbox{with}\quad
\normfact=
\frac{1}{\Gamma(D/2-1)}
\left(\frac{4\pi\muD^2}{-s}\right)^{2-D/2},
\eeqar
where $\alpha=e^2/(4\pi)$ is the electromagnetic 
fine-structure constant,
$\de \FF{\si}{l}$ represent the $l$-loop corrections relative 
to the tree-level form factors $\FF{\si}{0}$,
and  the normalization factor $\normfact$, 
with $\muD$ being the scale of dimensional regularization,
is introduced in
order to facilitate the comparison with 
\citere{Catani:1998bh}.
In the calculation, the $l$-loop 
contributions $\de \FF{\si}{l}$ are extracted from the corresponding 
vertex functions $F^\nu_{(l)}$
by means of the normalized projectors
\newcommand{\normprojector}[1]{\bar{\projector}^{(#1)}}
\beqar\label{normproject}
\normprojector{l}_\si
&=&
\frac{(4\pi)^{2l}}{\normfact^l}\,
\projector_\si,
\qquad
\normprojector{l}_\si
\left[F^\nu_{(l)}\right]
=
e^{2l} \FF{\si}{0}
\de \FF{\si}{l}.
\eeqar

The EW corrections 
depend on the invariant mass $s$, 
on the masses 
of the heavy virtual particles $\PW,\PZ,\Pt$, and $\PH$,
as well as on the renormalization scales  $\mu_i$.
In this paper we consider 
the asymptotic region
\beqar\label{asymptoticregion}
s\gg \MW^2\sim\MZ^2\sim\Mt^2\sim\MH^2,
\eeqar
where the invariant mass $s$ is much higher than
the EW scale.
In order to simplify the discussion,
let us assume for the moment that 
\beqar\label{equalmasses}
\MW=\MZ=\Mt=\MH,
\eeqar
and $\mu_i=\MW$.
In this case, the  EW corrections 
depend only on 
the dimensionless ratio $s/\MW^2$, and
are dominated by 
mass-singular logarithms%
\footnote{
These logarithms are evaluated 
in the Euclidean region $-s\gg \MW^2$,
where  the corrections are real. The imaginary parts 
that arise in the physical region
\refeq{asymptoticregion}
can be obtained via analytic continuation
replacing $s$ by $s+\ri 0$.
} 
\beqar\label{logsymbol}
L&:=&\log\left(\frac{-s}{\MW^2}\right),
\eeqar
which originate from soft and collinear heavy particles. 
In addition we have 
mass singularities 
that originate from soft and collinear massless photons
and give rise 
to $1/\veps$ poles  in $D=4-2 \veps$ dimensions.
Therefore,
the $l$-loop contributions to the chiral form factors can be written
as a double expansion in $L$ and $\veps$,
\beqar\label{logexpansion}
\de\FF{\si}{l}
=\sum_{j=0}^{2l}\sum_{k=-j}^{\infty}\de\FF{\si;j,k}{l}
\,\veps^{k}
L^{j+k}
,
\eeqar
with coefficients $\de \FF{\si;j,k}{l}$ that tend to constants
in the asymptotic limit $s/\MW^2\to\infty$.

In order to classify mass singularities,
we define the degree of singularity of a term in \refeq{logexpansion}
as the total power $j$ of logarithms $L$ and $1/\veps$ poles.
The maximal degree of singularity  at $l$-loop level is $j=2l$
and the corresponding terms are denoted as LL's.
The terms with $j= 2l-1$ represent the NLL's, and so on.

In this paper we 
systematically neglect mass-suppressed corrections of order $\MW^2/s$
and we calculate  the
one- and two-loop corrections to NLL accuracy,
\ie including LL's and NLL's.
For this approximation we use the symbol $\NLLA$.
The two-loop corrections are expanded 
in $\veps$ up to the finite terms, \ie contributions 
of order  $\veps^0 L^4$ and $\veps^0 L^3$.
Instead, the one-loop corrections are expanded up to order
$\veps^2$, \ie including terms of order 
 $\veps^2 L^4$ and $\veps^2 L^3$. As we will
see, these higher-order terms in the $\veps$-expansion
must be taken into account 
in the relation between one- and two-loop 
mass singularities.

In general, we have also logarithms
 \beqar
\Lmui&:=&\log\left(\frac{\mu_i^2}{\MW^2}\right),
\eeqar
which  depend on the 
renormalization scales  
$\mu_i=\muR,\mue$
of the weak mixing angle and the electromagnetic coupling constant.
In the expansion \refeq{logexpansion},
such logarithms are counted with the same weight as $L$,
\ie the terms of type $\veps^k\Lmui^m L^{j+k-m}$ 
are considered to be of the same order as $\veps^k L^{j+k}$.
Moreover, since the heavy-particle masses are of the same order,
\beqar\label{simmasses}
\MW\sim\MZ\sim\Mt\sim\MH,
\eeqar
but not equal, the coefficients 
$\de \FF{\si;j,k}{l}$ in \refeq{logexpansion}
depend also on the ratios of these masses\footnote{
This dependence starts at the subleading level, \ie for $j\le 2l-1$.
}.
Actually, as we will see in the final result, 
they depend only on logarithms of the type 
\beqar
\lr{i}:=\log\left(\frac{M_i^2}{\MW^2}\right)
.
\eeqar
These logarithms  are small compared to \refeq{logsymbol}, and 
in principle one could neglect them,
\ie one could regard the equal-mass case \refeq{equalmasses},
where they vanish,
 as a good approximation of the final result.
However, we find that 
this equal-mass approximation 
cannot be adopted in order 
to simplify
the analytical evaluation of the
two-loop diagrams.
In fact, as we will 
discuss in \refse{se:ewresult},
certain two-loop integrals
yield  contributions that depend not logarithmically 
but linearly on the ratio $\MW^2/\MZ^2$.
In this case, the relation 
between the weak-boson masses and the weak mixing angle
must be carefully taken into
account in order not to destroy cancellations that guarantee
the correct infrared behaviour of the two-loop corrections.
This means that
one cannot perform the diagrammatic calculation
by using the same mass in all 
heavy-particle propagators.

\subsection{Gauge interactions, symmetry breaking and mixing}
\label{se:gaugeint}
For the Feynman rules
we adopt the formalism of \citere{Pozzorini:rs}
(see App.~B),
which has been introduced 
in order to facilitate calculations 
in the high-energy limit of the spontaneously broken EW theory
and permits, in particular, to make extensive use of
various group-theoretical identities that can be found in
App.~A of \citere{Pozzorini:rs}.
In order to make the reader familiar with this 
formalism, we briefly review 
those aspects of spontaneous  symmetry breaking
that play a crucial role in the present calculation.

Within the unbroken phase of the EW theory, the gauge interactions are generated  by the covariant derivative
\beqar\label{covariantderivative1}
\mathcal{D}_\mu&=&
\partial_\mu-\ri e \sum_{\tilde{V}=W^1,W^2,W^3,B}
\tilde{V}_\mu \tilde{I}^{\tilde{V}}
,
\eeqar
where the vector
bosons $W^1,W^2,W^3$ and $B$ are associated to the 
generators
\beqar\label{unbrokengenerators}
e \tilde{I}^{W^a} = \gw T^a
,\qquad
e \tilde{I}^{B} = - \gb \frac{Y}{2},
\eeqar
of the $\SUtwo\times\Uone$ symmetry group
with the corresponding coupling constants $\gw$ and $\gb$.
The symmetry is spontaneously broken by a scalar Higgs doublet%
\beqar\label{Higgsdoublet}
\Phi=
\left(
\begin{array}{c}
\phi^+
\\
\frac{1}{\sqrt{2}}(\vev+H+\ri \chi)
\end{array}
\right)
\eeqar
which acquires a vacuum expectation value (vev) $\vev$ corresponding to the
minimum of its potential.
The Higgs doublet is parametrized in terms
of the four degrees of freedom 
$\Phi_i=\phi^+,\phi^-,H,\chi$, where 
$\phi^-=(\phi^+)^+$. In this representation the gauge-group  generators 
are $4\times 4$ matrices with components
$\tilde{I}^{\tilde{V}}_{\Phi_i\Phi_j}$.
The interactions of the gauge bosons with the vev generate the vector-boson mass matrix
\beqar\label{massmatrix}
M^2_{\tilde{V}\tilde{V}'}=\frac{1}{2}e^2 \vev^2 \left\{
\tilde{I}^{\tilde{V}},
\tilde{I}^{\tilde{V}'}
\right\}_{HH},
\eeqar
where the curly brackets denote an anticommutator.
The physical gauge bosons  A, Z, and $\PWpm$,
are the mass eigenstates of this matrix, which  
result from the unitary
transformation
\beqar\label{gbmixing}
W^\pm_\mu&=&\frac{1}{\sqrt{2}}\left(W_\mu^1\mp\ri W_\mu^2\right)
,\quad
Z_\mu=\cw W^3_\mu+\sw B_\mu 
,\quad
A_\mu=-\sw W^3_\mu+\cw B_\mu 
,
\eeqar
where 
$\cw=\cos\theta_\mathrm{w}$,
$\sw=\sin\theta_\mathrm{w}$,
and $\theta_\mathrm{w}$ is the weak mixing angle.
Their  masses are 
\beqar\label{massspectrum}
M_{\PWpm}&=&\frac{1}{2} \gw \vev
,\qquad
\MZ=\frac{1}{2\cw} \gw \vev
,\qquad
M_A=0,
\eeqar
and $\theta_\mathrm{w}$
is related to the weak-boson masses  via
\beqar\label{massrelation}
\MW=\cw \MZ.
\eeqar
The vanishing mass of the photon is connected to the fact that the 
electric charge of the vev is zero. 
This provides the relation
\beqar\label{neutralhiggs}
\cw \gb Y_\Phi =\sw \gw,
\eeqar
between the weak mixing angle, the coupling constants $\gb$, $\gw$ 
and the hypercharge $Y_\Phi$ of the Higgs doublet.
In the calculation 
we keep $Y_\Phi$ as a free parameter, which determines the degree of mixing in the gauge sector.
This permits us 
to consider the Standard Model case, $Y_\Phi=1$, as well as the special case  $Y_\Phi=0$ corresponding to  an unmixed theory with
\beqar\label{unmixed}
\sw=0
,\qquad
\cw=1
,\qquad
Z_\mu=W^3_\mu
,\qquad
A_\mu=B_\mu
,\qquad
\MW=\MZ.
\eeqar

In general, 
the mass-eigenstate gauge bosons 
are associated to the 
generators
\beqar\label{genmixing}
e {I}^{\pm} &=& \frac{\gw }{\sqrt{2}}\left(T^1\pm\ri T^2\right)
,\quad
e I^Z=
\cw \gw T^3- \sw \gb \frac{Y}{2}
,\quad
e I^A=
-\sw \gw T^3-\cw \gb\frac{Y}{2} 
\nln
\eeqar
with commutation relations
\beqar\label{commrel}
e \left[I^{V_1},
I^{V_2}
\right]
=\ri \gw \sum_{V_3=A,Z,W^\pm}
\teps^{V_1V_2V_3}
I^{\bar{V}_3},
\eeqar
where $\bar{V}$ denotes the complex conjugate of $V$,
\beq\label{phystotalantitens}
\varepsilon^{V_1 V_2V_3}=
\ri \times \left\{\begin{array}{c@{\quad}l} 
 (-1)^{p+1} \,\cw & \mbox{if}\quad V_1V_2V_3= \pi(Z W^+W^-), \\  
 (-1)^p \,\sw & \mbox{if}\quad V_1V_2V_3= \pi(A W^+W^-), \\  
0 & \mbox{otherwise,} 
\end{array}\right.
\eeq
and $(-1)^p$ represents the sign of the permutation $\pi$.
It is also useful to introduce 
the  $4\times 4$ matrices 
\beq\label{adjKron2}
\detwo=\left(\begin{array}{c@{\;}c@{\;}c@{\quad}c}
\sw^2 & -\sw\cw & 0 & 0 \\ 
-\sw\cw & \cw^2 & 0 & 0 \\
0 & 0 & 1 & 0 \\
0 & 0 & 0 & 1
\end{array}\right),
\qquad
\deone=\left(\begin{array}{c@{\;}c@{\;}c@{\quad}c}
\cw^2 & \sw\cw & 0 & 0 \\ 
\sw\cw & \sw^2 & 0 & 0 \\
0 & 0 & 0 & 0 \\
0 & 0 & 0 & 0
\end{array}\right),
\eeq
which 
project on the SU(2) and U(1)
components of the gauge fields, respectively%
\footnote{
The order of the components 
$\detwo_{{V}V'}$ and $\deone_{{V}V'}$
in \refeq{adjKron2}
corresponds to $V, V'=A,Z,W^+,W^-$. 
For instance we can project the SU(2) and U(1) contributions to the
generators with 
\beqar\label{weakprojector1}
e \sum_{V=A,Z,W^\pm} 
\detwo_{A{V}} I^{\bar{V}}&=&
-\gw \sw T^3
,\qquad
e \sum_{V=A,Z,W^\pm} 
\detwo_{Z{V}} I^{\bar{V}}=
\gw \cw T^3
,\nn
\eeqar
and
\beqar\label{weakprojector2}
e \sum_{V=A,Z,W^\pm} 
\deone_{A{V}} I^{\bar{V}}&=&
-\gb \cw \frac{Y}{2}
,\qquad
e \sum_{V=A,Z,W^\pm} 
\deone_{Z{V}} I^{\bar{V}}=
-\gb \sw \frac{Y}{2}
.\nn
\eeqar
}.
Analogously to the SU(2) structure constants,
the $\teps$ tensor \refeq{phystotalantitens}
satisfies the identities
\beqar
\sum_{V_3=A,Z,W^\pm} \teps^{\bar{V}_1 \bar{V}_2 \bar{V}_3}\teps^{V'_{1 } V'_{2}{V}_3}
&=&\detwo_{{V}_{1}V'_{1}}\detwo_{{V}_{2}V'_{2}}-\detwo_{{V}_{1}V'_{2}}\detwo_{{V}_{2}V'_{1}},\nl
\sum_{V_2,V_3=A,Z,W^\pm} \teps^{\bar{V}_1 \bar{V}_2 \bar{V}_3}\teps^{V'_{1} {V}_{2}{V}_3}
&=&\CA \detwo_{{V}_{1}V'_{1}},
\eeqar
where  $\CA=2$ is the eigenvalue of the 
Casimir operator in the adjoint representation.
The dimension of the $\SUtwo$ gauge group is denoted as 
$\dimtwo=\Tr[\detwo]=3$.

The above 
group-theoretical identities have been used in the calculation
in order to express all combinations of (non-commuting) 
fermionic chiral couplings $I^V_\si$ occurring in Feynman diagrams 
in terms of the 
electric charge 
$Q=-I^A$, the 
diagonal matrices $Y_\si,T^3_\si$, and  
 the SU(2) Casimir operator $C_\si$, 
with $C_+=0$ and $C_-=3/4$.

\subsection{Mass-gap effects}
\label{se:massgapeff}
The EW corrections 
depend on various mass scales: 
the masses of the heavy virtual particles,
which we generically denote as 
$M_i=\MW,\MZ,\Mt,\MH$,
and in general also a photon mass $\la$, 
which can be used to regularize mass singularities 
from virtual photons,
\beqar\label{subtractiondef0}
\de \FF{\si}{l}&\equiv&
\de \FF{\si}{l}\left(\{M_i\},\la\right)
.
\eeqar
In our calculation the photon and the light fermions are
massless, 
and 
the heavy masses  \refeq{simmasses}
are assumed to be of the same order but not equal.
One of our aims is to investigate the effects resulting 
from  the splittings between these masses, and in particular
from the gap between the photon mass ($\la=0$)
and the weak-boson masses,
which is a non-trivial aspect of the EW corrections.
To this end, it is useful to isolate the mass-gap effects 
by adding and subtracting an equal-mass contribution, 
where the EW corrections 
are evaluated by setting  $M_i=\MW$ and $\la=\MW$
in the propagators of the heavy particles and
of the photon.
This results in the splitting  
\beqar\label{subtractiondef}
\de \FF{\si}{l}\left(\{M_i\},0\right)
&=&
\left.\de \FF{\si}{l}\left(\{M_i\},\MW\right)\right|_{M_i=\MW}+
\Delta\,\de \FF{\si}{l}\left(\{M_i\},\pmass\right),
\eeqar
where all  mass-gap effects, and in particular all
infrared divergences 
(or combinations of infrared and ultraviolet divergences),
are isolated in the subtracted term
$\Delta\,\de \FF{\si}{l}$,
whereas the purely ultraviolet divergences are contained in 
the equal-mass term.
According to the IREE, 
the equal-mass term is expected to behave as in a 
symmetric $\SUtwo\times\Uone$ theory, 
whereas the mass-gap term,
or more precisely the part resulting 
from the gap between the photon and the weak-boson masses, 
is expected to be of QED nature.

The analytical results for one- and two-loop diagrams
that we present in the next sections 
are systematically split as in 
\refeq{subtractiondef}.
The relevant loop integrals
$\DD{i}$ and corresponding subtracted 
functions $\deDD{i}$ are defined in \refapp{app:loops}.
In all diagrams,
the photonic couplings which occur in the equal-mass contributions 
are always written in terms of $Y$ and $T^3$, 
whereas in the mass-gap contributions we always write $Q$.
This permits us, at the end of the calculation, 
to switch off all contributions 
originating from the $\PA-\PW$ mass gap,
\ie to set $\la=\MW$,
by simply substituting $Q=0$.

The integration of the massive 
one- and two-loop integrals 
has been  performed 
to NLL accuracy by means of an automatized algorithm based on the 
so-called sector-decomposition technique 
\cite{Hepp:1966eg,Roth:1996pd,Binoth:2000ps,Denner:2002gd}.
This method 
permits
to separate 
overlapping ultraviolet 
and mass singularities in 
Feynman-parameter integrals.
As a result of this sector decomposition, 
the integrand can be 
split into 
simple subtraction terms 
that lead to ultraviolet and mass singularities
and a finite remainder. 
Finally, the  mass-singular integration
of the subtraction terms that lead to the LL and NLL contributions
can be performed in analytic form.
A detailed description of the sector-decomposition algorithm 
for massive loop integrals is postponed to a forthcoming paper 
\cite{nextpaper}.

\section{One-loop form factors}
\label{se:oneloop}
In this section we present the one-loop contributions
to the bare and renormalized form factors
\refeq{projectordef}.
Each result is expanded in $\veps$ up to the order $\veps^2$.

\subsection{Bare form factors}
At one loop only the following topology contributes
\beqar
\normprojector{1}_\si\quad
\vcenter{\hbox{
\diagone
}}
\equaldiag
e^2 
\FF{\si}{0}
\sum_{a_1=A,Z, \pm
} 
I_\si^{\bar{a}_1}
I_\si^{{a}_1}
\DD{0}(M_{a_1}).
\eeqar
The corresponding loop integral $\DD{0}$ 
is defined in \refeq{defint0} and to NLL accuracy 
yields
\beqar\label{idiag0}
\DD{0}(\MW)&\NLLA& \ResoneM,
\nl
\deDD{0}(\MZ)&\NLLA& 
\left(\ResdeoneM
\right)
\LMZW
,
\nl
\deDD{0}(\pmass)&\NLLA& \ResdeoneZ.
\eeqar
This leads to
\beqar\label{cdiag0}
e^2 \de \FF{\si,0}{1}&=&
\coeffsymbol_{\si,0}^{\sew}
\DD{0}(\MW)
+
\coeffsymbol_{\si,0}^{Z}
\deDD{0}(\MZ)
+
\coeffsymbol_{\si,0}^{A}
\deDD{0}(\pmass),
\eeqar
where
\beqar\label{c2diag0}
\coeffsymbol_{\si,0}^{\sew}
&=&
e^2\sum_{a_1=A,Z,\pm
} 
I_\si^{\bar{a}_1}
I_\si^{{a}_1}
=\gb^2\left(\frac{Y_\si}{2}\right)^2+\gw^2\ctwo
,\nl
\coeffsymbol_{\si,0}^{Z}
&=&
e^2I_\si^{Z}I_\si^{Z}
=\gb^2\left(\frac{Y_\si}{2}\right)^2 +\gw^2(T_\si^3)^2
-e^2 Q^2
,\qquad
\coeffsymbol_{\si,0}^{A}
=
e^2 Q^2.
\eeqar

\subsection{Renormalization}
The tree-level form factors 
$\FF{\si}{0}$
do not depend on 
the EW coupling constants
and masses. Therefore, 
the only one-loop counterterm contribution arises 
from the on-shell 
wave-function renormalization constants (WFRC's) 
for massless chiral fermions%
\footnote{
Note that we adopt a perturbative expansion 
with the same 
normalization factor $\normfact$ as in \refeq{pertserie1}.
},
\beqar\label{WFpertserie1}
Z_\si
&=&
1+\sum_{l=1}^\infty
\left(\frac{\alpha}{4\pi}\right)^l
\normfact^l
\de Z^{(l)}_\si,
\eeqar
which yield
\beqar\label{WFren}
\de^{\mathrm{WF}} 
\FF{\si}{1}&=&
\de Z^{(1)}_{\si} 
.
\eeqar
The one-loop WFRC's receive contributions only from massive
weak bosons, whereas the photonic contribution vanishes 
owing to a cancellation between ultraviolet
and mass singularities within dimensional regularization.
To NLL accuracy we have
\beqar\label{WFRC}
e^2 \de Z^{(1)}_{\si}&\NLLA&
-
\frac{1}{\veps}
\sum_{a\neq A} 
e^2I_\si^{\bar{a}}
I_\si^{{a}}\,
\left(\frac{-s}{M_a^2}\right)^\veps
\\&=&
-
\frac{1}{\veps}
\left\{
\gw^2\left[\ctwo -(T_\si^3)^2\right]
\left(\frac{-s}{\MW^2}\right)^\veps
{}+
\left[\gb^2\left(\frac{Y_\si}{2}\right)^2 +\gw^2(T_\si^3)^2
-e^2 Q^2
\right]
\left(\frac{-s}{\MZ^2}\right)^\veps
\right\}
\nn.
\eeqar

\subsection{Renormalized one-loop form factors}
\label{se:oneloopres}
Combining the above results
we obtain the  renormalized 
one-loop form factors 
\beqar\label{onelooprena}
e^2 
\de \FF{\si}{1}
&=&
e^2 \left[\de \FF{\si,0}{1}
+ 
\de^{\mathrm{WF}} \FF{\si}{1}
\right]
\nl&\NLLA&
\coeffsymbol_{\si,0}^{\sew}
\univfact{\veps}{\MW}
+
\coeffsymbol_{\si,0}^{Z}
\Delta
\univfact{\veps}{\MZ}
+
\coeffsymbol_{\si,0}^{A}
\Delta\univfact{\veps}{0}
,
\eeqar
where expanding in  $\veps$ 
up to the order $\veps^2$ we have
\beqar\label{Ifunc}
\univfact{\veps}{\MW}&\NLLA&
\ResIunivW
+\order(\veps^3)
,\nl
\univfact{\veps}{\MZ}&\NLLA&
\univfact{\veps}{\MW}+
\LMZW\left(\ResdeIunivZ\right)
+\order(\veps^3)
,\nl
\univfact{\veps}{0}&\NLLA&
\ResIunivP
,
\eeqar
and the subtracted functions $\Delta I$ are defined as
\beqar\label{subtractionofI}
\Delta\univfact{\veps}{m}&:=&
\univfact{\veps}{m}
-\univfact{\veps}{\MW}.
\eeqar

\section{Two-loop form factors}
\label{se:twoloop}
In this section we present
the  two-loop contributions from individual Feynman diagrams 
and counterterms.
Each result is expanded in $\veps$ up to the order $\veps^0$.
The diagrams 1--3 in \refse{se:baretwoloop} give rise to LL's
and NLL's, whereas all other diagrams and counterterm contributions 
yield only NLL's.

\subsection{Bare form factors}
\label{se:baretwoloop}
The two-loop Feynman diagrams contributing to the bare form factors 
involve virtual gauge bosons ($a_i=\PA,\PZ,\PWpm$, $\bar{a}_i=\PA,\PZ,\PW^\mp$),  
ghosts ($u^a_i=u^A,u^\PZ,u^\PWpm$),
Higgs bosons and would-be Goldstone bosons ($\Phi_i=H,\chi,\phi^\pm$),
as well as fermionic doublets ($\Psi_i=u,d$). 
The corresponding loop integrals are defined in \refapp{app:loops}.
The computation is performed within the 
 't~Hooft--Feynman gauge, where the masses of
the ghosts and would-be-Goldstone bosons read
$M_{\phi^\pm}=M_{u^\pm}=\MW$ 
and $M_{\chi}=M_{u^Z}=\MZ$. 
The symbol $M_i$ 
is used to denote generic heavy masses, 
\ie  $M_i=\MW,\MZ,\Mt,\MH$.

\subsubsection{Diagram \one}
\beqar\label{diagram1}
\normprojector{2}_\si\quad
\vcenter{\hbox{
\diagI
}}
\equaldiag
e^4  \FF{\si}{0}
\sum_{a_1,a_2=A,Z,\pm} 
I_\si^{\bar{a}_1}
I_\si^{\bar{a}_2}
I_\si^{{a}_2}
I_\si^{{a}_1}
\DD{1}(M_{a_1},M_{a_2})
,
\eeqar
where the loop integral $\DD{\one}$ is defined in \refeq{defint1} and yields
\beqar\label{idiag1}
\DD{\one}(\MW,\MW)&\NLLA& \ResIMM,
\nl
\deDD{\one}(M_1,M_2)&\NLLA& \ResdeIMM
\LrMI
,
\nl
\deDD{\one}(\pmass,M_2)&\NLLA& 
\ResdeIZM,
\nl
\deDD{\one}(M_1,\pmass)&\NLLA& \ResdeIMZ
\LrMI
,\nl
\deDD{\one}(\pmass,\pmass)&\NLLA& \ResdeIZZ.
\eeqar
This leads to
\beqar\label{cdiag1}
e^4 \de \FF{\si,\one}{2}&\NLLA&
\coeffsymbol_{\si,\one}^{\sew}
\DD{\one}(\MW,\MW)
+
\coeffsymbol_{\si,\one}^{Z}
\deDD{\one}(\MZ,\MW)
+
\coeffsymbol_{\si,\one}^{A W}
\deDD{\one}(\pmass,\MW)
\nl&&{}+
\coeffsymbol_{\si,\one}^{A Z}
\deDD{\one}(\pmass,\MZ)
+
\coeffsymbol_{\si,\one}^{A A}
\deDD{\one}(\pmass,\pmass)
,
\eeqar
where
\beqar\label{c2diag1}
\coeffsymbol_{\si,\one}^{\sew}
&=& e^4
\sum_{a_1,a_2
=A,Z,\pm
} 
I_\si^{\bar{a}_1}
I_\si^{\bar{a}_2}
I_\si^{{a}_2}
I_\si^{{a}_1}
=\left[\gb^2\left(\frac{Y_\si}{2}\right)^2+\gw^2\ctwo\right]^2
,
\nl
\coeffsymbol_{\si,\one}^{Z}
&=&
e^4
\sum_{a_2
=A,Z,\pm} 
I_\si^{Z}
I_\si^{\bar{a}_2}
I_\si^{{a}_2}
I_\si^{Z}
=
e^2 \left(I_\si^Z\right)^2
\left[\gb^2\left(\frac{Y_\si}{2}\right)^2+\gw^2\ctwo\right]
,
\nl
\coeffsymbol_{\si,\one}^{A W}
&=&
e^4
\sum_{a_2=\pm} 
I_\si^{A}
I_\si^{\bar{a}_2}
I_\si^{{a}_2}
I_\si^{A}
=e^2 \gw^2 Q^2\left[\ctwo -(T_\si^3)^2\right]
,
\nl
\coeffsymbol_{\si,\one}^{A Z}
&=&
e^4 Q^2 \left(I_\si^Z\right)^2
,
\qquad
\coeffsymbol_{\si,\one}^{A A}
=
e^4 Q^4.
\eeqar
Note that  
all $\deDD{}$-contributions 
from diagrams with $a_1=Z,\pm$ and $a_2=A,Z,\pm$ 
have been combined in the term 
proportional to 
$\coeffsymbol_{\si,\one}^{Z}$ 
in \refeq{cdiag1}
using 
the relations
\beqar
\deDD{\one}(\MZ,\MW)
&\NLLA&
\deDD{\one}(\MZ,\MZ)
\NLLA
\deDD{\one}(\MZ,\pmass),
\nl
\deDD{\one}(\MW,\MW)
&\NLLA&
\deDD{\one}(\MW,\MZ)
\NLLA
\deDD{\one}(\MW,\pmass)\NLLA
0,
\eeqar
which can be read off from \refeq{idiag1}.
For the diagrams that we present in the following
similar simplifications are implicitly understood.

\subsubsection{Diagram \two}
\beqar\label{diagram2}
\normprojector{2}_\si\quad
\vcenter{\hbox{
\diagII
}}
\equaldiag
e^4  \FF{\si}{0}
\sum_{a_1,a_2
=A,Z,\pm} 
I_\si^{\bar{a}_1}
I_\si^{\bar{a}_2}
I_\si^{{a}_1}
I_\si^{{a}_2}
\DD{\two}(M_{a_1},M_{a_2})
,
\eeqar
where the loop integral $\DD{\two}$ is defined in \refeq{defint2} and yields
\beqar\label{idiag2}
\DD{\two}(\MW,\MW)&\NLLA& \ResIIMM,
\nl
\deDD{\two}(M_1,M_2)&\NLLA& \ResdeIIMM
\left(\LrMI+\LrMII\right)
,
\nl
 \deDD{\two}(\pmass,M_2)&\NLLA& \ResdeIIZM,
\nl
 \deDD{\two}(M_1,\pmass)&\NLLA& \ResdeIIMZ,
\nl
 \deDD{\two}(\pmass,\pmass)&\NLLA& \ResdeIIZZ.
\eeqar
This leads to
\beqar\label{cdiag2}
e^4 \de \FF{\si,\two}{2}&\NLLA&
\coeffsymbol_{\si,\two}^{\sew}
\DD{\two}(\MW,\MW)
+
\coeffsymbol_{\si,\two}^{WZ}
\deDD{\two}(\MW,\MZ)
+
\coeffsymbol_{\si,\two}^{W A }
\deDD{\two}(\MW,\pmass)
\nl&&{}+
\coeffsymbol_{\si,\two}^{ZZ}
\deDD{\two}(\MZ,\MZ)
+
\coeffsymbol_{\si,\two}^{ZA}
\deDD{\two}(\MZ,\pmass)
+
\coeffsymbol_{\si,\two}^{A A}
\deDD{\two}(\pmass,\pmass)
,
\eeqar
where
\beqar\label{c2diag2}
\coeffsymbol_{\si,\two}^{\sew}
&=&e^4
\sum_{a_1,a_2
=A,Z,\pm} 
I_\si^{\bar{a}_1}
I_\si^{\bar{a}_2}
I_\si^{{a}_1}
I_\si^{{a}_2}
=
\left[\gb^2\left(\frac{Y_\si}{2}\right)^2+\gw^2\ctwo\right]^2-\frac{1}{2}\gw^4 \CA \ctwo
,\nl
\coeffsymbol_{\si,\two}^{WZ}
&=&e^4
\sum_{a_1=\pm} 
\left(
I_\si^{\bar{a}_1}
I_\si^{Z}
I_\si^{{a}_1}
I_\si^{Z}+
I_\si^{Z}
I_\si^{\bar{a}_1}
I_\si^{Z}
I_\si^{{a}_1}
\right)=
2 e^2 \gw^2 \left(I_\si^Z\right)^2
\left[\ctwo- (T_\si^3)^2\right]
\nl&&{}-
e \gw^3\cw \CA T_\si^3 I_\si^Z
,\nl
\coeffsymbol_{\si,\two}^{WA}
&=&e^4
\sum_{a_1=\pm} 
\left(
I_\si^{\bar{a}_1}
I_\si^{A}
I_\si^{{a}_1}
I_\si^{A}+
I_\si^{A}
I_\si^{\bar{a}_1}
I_\si^{A}
I_\si^{{a}_1}
\right)
\nl&=&
2 e^2 \gw^2 Q^2
\left[\ctwo- (T_\si^3)^2\right]
-
e \gw^3\sw \CA T_\si^3 Q 
,\nl
\coeffsymbol_{\si,\two}^{ZZ}
&=&e^4 \left(I_\si^Z\right)^4
,\qquad
\coeffsymbol_{\si,\two}^{ZA}
=2 e^4 Q^2\left(I_\si^Z\right)^2
,\qquad
\coeffsymbol_{\si,\two}^{AA}
=e^4 Q^4.
\eeqar

\subsubsection{Diagrams \thr}
\beqar\label{diagram3}
&&\normprojector{2}_\si\quad
\vcenter{\hbox{
\diagIII
}}
+\quad
\normprojector{2}_\si\quad
\vcenter{\hbox{
\diagIV
}}
\equaldiag
\nl&&\qquad{}=
-2\ri e^3
\gw
\FF{\si}{0}
\sum_{a_1,a_2,a_3
=A,Z,\pm} 
\teps^{a_1 a_2 a_3}
I_\si^{\bar{a}_1}
I_\si^{\bar{a}_2}
I_\si^{\bar{a}_3}
\DD{\thr}(M_{a_1},M_{a_2},M_{a_3})
,
\eeqar
where the $\teps$ tensor is defined in \refeq{phystotalantitens}.
The loop integral $\DD{\thr}$ is defined in \refeq{defint3} and yields
\beqar\label{idiag3}
\DD{\thr}(\MW,\MW,\MW)&\NLLA&\ResIIIMMM,
\nl
\deDD{\thr}(M_1,M_2,M_3)&\NLLA&\ResdeIIIMMM
\left(\LrMI+\LrMIII\right)
,
\nl
 \deDD{\thr}(\pmass,M_2,M_3)&\NLLA&\ResdeIIIZMM,
\nl
 \deDD{\thr}(M_1,\pmass,M_3)&\NLLA&\ResdeIIIMZM
\left(\LrMI+\LrMIII\right)
,
\nl
\deDD{\thr}(M_1,M_2,\pmass)&\NLLA& \ResdeIIIMMZ
.
\eeqar
This leads to
\beqar\label{cdiag3}
e^4 \de \FF{\si,\thr}{2}&=&
\coeffsymbol_{\si,\thr}^{\sew}
\DD{\thr}(\MW,\MW,\MW)
+
\coeffsymbol_{\si,\thr}^{A WW}
\deDD{\thr}(\pmass,\MW,\MW)
\nl&&{}
+
\coeffsymbol_{\si,\thr}^{WWA}
\deDD{\thr}(\MW,\MW,\pmass)
+
\coeffsymbol_{\si,\thr}^{ZWW}
\deDD{\thr}(\MZ,\MW,\MW)
,
\eeqar
where
\beqar\label{c2diag3}
\coeffsymbol_{\si,\thr}^{\sew}
&=&
-2\ri e^3 \gw
\sum_{a_1,a_2,a_3
=A,Z,\pm} 
\teps^{a_1 a_2 a_3}
I_\si^{\bar{a}_1}
I_\si^{\bar{a}_2}
I_\si^{\bar{a}_3}
=
\gw^4 \CA \ctwo
,\nl
\coeffsymbol_{\si,\thr}^{A WW}
&=&
-2\ri e^3 \gw
\sum_{a_2,a_3=\pm} 
\teps^{A a_2 a_3}
I_\si^{A}
I_\si^{\bar{a_2}}
I_\si^{\bar{a}_3}
=
e \gw^3 \sw \CA Q T_\si^3
,\nl
\coeffsymbol_{\si,\thr}^{WWA}
&=&
-2\ri e^3 \gw
\sum_{a_1,a_2=\pm} 
\teps^{a_1 a_2 A}
I_\si^{\bar{a_1}}
I_\si^{\bar{a}_2}
I_\si^{A}
=
\coeffsymbol_{\si,\thr}^{A WW}
,\nl
\coeffsymbol_{\si,\thr}^{ZWW}
&=&
-2\ri e^3 \gw
\left(\sum_{a_2,a_3=\pm} 
\teps^{Z a_2 a_3}
I_\si^{Z}
I_\si^{\bar{a_2}}
I_\si^{\bar{a}_3}
+
\sum_{a_1,a_2=\pm} 
\teps^{a_1 a_2 Z}
I_\si^{\bar{a_1}}
I_\si^{\bar{a}_2}
I_\si^{Z}
\right)
\nl&=&
2 e \gw^3 \cw \CA I_\si^Z T_\si^3
.
\eeqar

\subsubsection{Diagrams \fiv}
\beqar\label{diagram5}
&&\normprojector{2}_\si\quad
\vcenter{\hbox{
\diagV
}}
+\quad
\normprojector{2}_\si\quad
\vcenter{\hbox{
\diagVI
}}
\equaldiag
\nl&&\qquad{}=
2 e^4
 \FF{\si}{0}
\sum_{a_1,a_2
=A,Z,\pm} 
I_\si^{\bar{a}_1}
I_\si^{\bar{a}_2}
I_\si^{{a}_2}
I_\si^{{a}_1}
\DD{\fiv}(M_{a_1},M_{a_2})
,
\eeqar
where the loop integral $\DD{\fiv}$ is defined in \refeq{defint5} and yields
\beqar\label{idiag5}
\DD{\fiv}(\MW,\MW)&\NLLA& \ResVMM,
\nl
\deDD{\fiv}(M_1,M_2)&\NLLA& \ResdeVMM,
\nl
 \deDD{\fiv}(\pmass,M_2)&\NLLA& \ResdeVZM,
\nl
 \deDD{\fiv}(M_1,\pmass)&\NLLA& \ResdeVMZ,
\nl
 \deDD{\fiv}(\pmass,\pmass)&\NLLA& \ResdeVZZ.
\eeqar
This leads to 
\beqar\label{cdiag5}
e^4 \de \FF{\si,\fiv}{2}&=&
\coeffsymbol_{\si,\fiv}^{\sew}
\DD{\fiv}(\MW,\MW)
+
\coeffsymbol_{\si,\fiv}^{A \wbos}
\deDD{\fiv}(\pmass,\MW)
+\coeffsymbol_{\si,\fiv}^{A A}
\deDD{\fiv}(\pmass,\pmass)
,
\eeqar
where 
\beqar\label{c2diag5}
\coeffsymbol_{\si,\fiv}^{\sew}
&=&2 e^4 \sum_{a_1,a_2
=A,Z,\pm} 
I_\si^{\bar{a}_1}
I_\si^{\bar{a}_2}
I_\si^{{a}_2}
I_\si^{{a}_1}
=2\left[\gb^2\left(\frac{Y_\si}{2}\right)^2+\gw^2\ctwo\right]^2
,
\nl
\coeffsymbol_{\si,\fiv}^{A \wbos}
&=&
2 e^4 \sum_{a_2\neq A} 
I_\si^{ A}
I_\si^{\bar{a}_2}
I_\si^{{a}_2}
I_\si^{ A}
=2e^2 Q^2\left[\gb^2\left(\frac{Y_\si}{2}\right)^2+\gw^2\ctwo- e^2 Q^2\right]
,
\nl
\coeffsymbol_{\si,\fiv}^{ A  A}
&=&
2e^4 Q^4.
\eeqar

\subsubsection{Diagrams \sev}
\beqar\label{diagram7}
&&\normprojector{2}_\si\quad
\vcenter{\hbox{
\diagVII
}}
+\quad
\normprojector{2}_\si\quad
\vcenter{\hbox{
\diagVIII
}}
\equaldiag
\nl&&\qquad{}=
2 e^4
 \FF{\si}{0}
\sum_{a_1,a_2
= A,Z,\pm} 
I_\si^{\bar{a}_2}
I_\si^{\bar{a}_1}
I_\si^{{a}_2}
I_\si^{{a}_1}
\DD{\sev}(M_{a_1},M_{a_2})
,
\eeqar
where the loop integral $\DD{\sev}$ is defined in \refeq{defint7} and yields
\beqar\label{idiag7}
\DD{\sev}(\MW,\MW)&\NLLA& \ResVIIMM,
\nl
\deDD{\sev}(M_1,M_2)&\NLLA& \ResdeVIIMM,
\nl
 \deDD{\sev}(\pmass,M_2)&\NLLA& \ResdeVIIZM,
\nl
 \deDD{\sev}(M_1,\pmass)&\NLLA& \ResdeVIIMZ,
\nl
 \deDD{\sev}(\pmass,\pmass)&\NLLA& \ResdeVIIZZ.
\eeqar
This leads to
\beqar\label{cdiag7}
e^4 \de \FF{\si,\sev}{2}&=&
\coeffsymbol_{\si,\sev}^{\sew}
\DD{\sev}(\MW,\MW)
+
\coeffsymbol_{\si,\sev}^{ A \wbos}
\deDD{\sev}(\pmass,\MW)
+\coeffsymbol_{\si,\sev}^{ A  A}
\deDD{\sev}(\pmass,\pmass)
,
\eeqar
where 
\beqar\label{c2diag7}
\coeffsymbol_{\si,\sev}^{\sew}
&=&2 e^4 \sum_{a_1,a_2
= A,Z,\pm} 
I_\si^{\bar{a}_1}
I_\si^{\bar{a}_2}
I_\si^{{a}_1}
I_\si^{{a}_2}
=
2\left[\gb^2\left(\frac{Y_\si}{2}\right)^2+\gw^2\ctwo\right]^2-\gw^4 \CA \ctwo
,\nl
\coeffsymbol_{\si,\sev}^{ A \wbos}
&=&2 e^4 \sum_{a_2\neq A} 
I_\si^{ A}
I_\si^{\bar{a}_2}
I_\si^{ A}
I_\si^{{a}_2}
=
2e^2 Q^2\left[\gb^2\left(\frac{Y_\si}{2}\right)^2+\gw^2\ctwo- e^2 Q^2\right]-
e \gw^3\sw \CA Q T_\si^3  
,\nl
\coeffsymbol_{\si,\sev}^{ A  A}
&=&
2e^4 Q^4.
\eeqar

\subsubsection{Diagrams \ten}
\label{sub:diagten}
\beqar\label{diagram10}
&&\normprojector{2}_\si\quad
\vcenter{\hbox{
\diagX
}}
+\quad
\normprojector{2}_\si\quad
\vcenter{\hbox{
\diagXI
}}
\equaldiag
\\&&\qquad{}=
-\frac{1}{2} e^2 \gw^2
 \FF{\si}{0}
\sum_{a_1,a_2,a_3,a_4
= A,Z,\pm} 
\teps^{a_1 \bar{a}_2 \bar{a}_3}
\teps^{a_4 a_2 a_3}\,
I_\si^{\bar{a}_1}
I_\si^{\bar{a}_4}\,
\DD{\ten}(M_{a_1},M_{a_2},M_{a_3},M_{a_4})
,\nn
\eeqar
where the loop integral $\DD{\ten}$ is defined in \refeq{defint10} and yields
\beqar\label{idiag10}
\DD{\ten}(\MW,\MW,\MW,\MW)
&\NLLA& \ResXMMMM,
\nl
\deDD{\ten}(M_1,M_2,M_3,M_4)
&\NLLA& \ResdeXMMMM,
\nl
\deDD{\ten}(\pmass,M_2,M_3,M_4)
&\NLLA& 
\left(\frac{M_2^2+M_3^2}{{2M_4^2}}\right)
\left[\ResdeXZMMM\right],
\nl
\deDD{\ten}(M_1,\pmass,M_3,M_4)
&\NLLA& 
\ResdeXMZMM,
\nl
\deDD{\ten}(M_1,M_2,\pmass,M_4)
&\NLLA& 
\ResdeXMMZM,
\nl
\deDD{\ten}(M_1,M_2,M_3,\pmass)
&\NLLA& 
\left(\frac{M_2^2+M_3^2}{{2M_1^2}}\right)
\left[\ResdeXMMMZ\right],
\nl
\deDD{\ten}(\pmass,M_2,M_3,\pmass)
&\NLLA& \ResdeXZMMZ.
\eeqar
This leads to
\beqar\label{cdiag10}
e^4 \de \FF{\si,\ten}{2}
&=&
\coeffsymbol_{\si,\ten}^{\sew}
\DD{\ten}(\MW,\MW,\MW,\MW)
+
\coeffsymbol_{\si,\ten}^{ A Z}
\deDD{\ten}(\pmass,\MW,\MW,\MZ)
\nl&&{}
+
\coeffsymbol_{\si,\ten}^{ A  A}
\deDD{\ten}(\pmass,\MW,\MW,\pmass),
\eeqar
where
\beqar\label{c2diag10}
\coeffsymbol_{\si,\ten}^{\sew}
&=&
-\frac{1}{2} e^2 \gw^2
\sum_{a_1,a_2,a_3,a_4
= A,Z,\pm} 
\teps^{a_1 \bar{a}_2 \bar{a}_3}
\teps^{a_4 a_2 a_3}\,
I_\si^{\bar{a}_1}
I_\si^{\bar{a}_4}
=
-\frac{1}{2}\gw^4 \CA \ctwo
,\nl
\coeffsymbol_{\si,\ten}^{ A Z}
&=&
-\frac{1}{2}  e^2 \gw^2
\left[I_\si^{ A}
I_\si^{Z}
\sum_{a_2,a_3=\pm} 
\teps^{ A \bar{a}_2\bar{a}_3}
\teps^{Z a_2 a_3}
+(A\leftrightarrow Z)\right]
=
-e^2 \gw^2 \sw\cw 
\CA Q I_\si^Z 
,\nl
\coeffsymbol_{\si,\ten}^{ A A}
&=&
-\frac{1}{2}  e^2 \gw^2
\sum_{a_2,a_3\neq A} 
\teps^{ A \bar{a}_2 \bar{a}_3}
\teps^{ A a_2 a_3}\,
I_\si^{ A}
I_\si^{ A}
=-\frac{1}{2}e^2 \gw^2\sw^2  \CA Q^2.
\eeqar
We observe that the loop integrals associated with 
A--Z mixing-energy subdiagrams
give rise to the contributions
\beqar\label{lineardep}
\deDD{\ten}(\pmass,\MW,\MW,\MZ)
&=&
\left(\frac{\MW}{{\MZ}}\right)^2
\left[\ResdeXZMMM\right],
\eeqar
which depend linearly on the ratio
$\left(\MW/\MZ\right)^2$.
This dependence appears
also in all other bosonic A--Z mixing-energy diagrams 
\sixteen--\seventeen~and  is discussed in detail 
in \refse{se:ewresult}. 
There we explain its origin and show that,
in order not to destroy cancellations
that ensure the correct infrared behaviour of the 
corrections,
the relation 
\refeq{massrelation}
between the weak-boson 
masses and the weak mixing angle
must be 
carefully taken  into account.
This means that 
such loop integrals 
cannot be evaluated using
$\MW=\MZ$  in the  weak-boson 
propagators.

\subsubsection{Diagram \sixteen}
\beqar\label{diagram16}
\normprojector{2}_\si\quad
\vcenter{\hbox{
\diagXVI
}}
\equaldiag
\begin{array}{l}
\\
\displaystyle{
e^2 \gw^2
 \FF{\si}{0}
\sum_{a_1,a_2,a_3
= A,Z,\pm} 
I_\si^{\bar{a}_1}
I_\si^{\bar{a}_3}
\left[
\de^{\SUtwo}_{\bar{a}_1a_3}
\de^{\SUtwo}_{\bar{a}_2 a_2}
-
\de^{\SUtwo}_{\bar{a}_1a_2}
\de^{\SUtwo}_{\bar{a}_2a_3}
\right]
}
\\
\displaystyle{\quad{}\times
(D-1) 
\DD{\sixteen}(M_{a_1},M_{a_2},M_{a_3})
,}
\end{array}
\nln
\eeqar
where $\de^{\SUtwo}$ is the matrix is defined in
\refeq{adjKron2}
and  $D=4-2\veps$.
The loop integral $\DD{\sixteen}$ is defined in \refeq{defint16} and yields
\beqar\label{idiag16}
\DD{\sixteen}(\MW,\MW,\MW)
&\NLLA&
\ResXVIIMMM
,\nl
\deDD{\sixteen}(M_1,M_2,M_3)
&\NLLA&
\ResdeXVIIMMM
,\nl
\deDD{\sixteen}(\pmass,M_2,M_3)
&\NLLA&
\left(\frac{M_2}{{M_3}}\right)^2
\left[\ResdeXVIIZMM\right]
,\nl
\deDD{\sixteen}(M_1,\pmass,M_3)
&\NLLA&
0
,\nl
\deDD{\sixteen}(M_1,M_2,\pmass)
&\NLLA&
\left(\frac{M_2}{{M_1}}\right)^2
\left[\ResdeXVIIMMZ\right],
\nl
\deDD{\sixteen}(\pmass,M_2,\pmass)
&\NLLA&
\ResdeXVIIZMZ.
\eeqar
This leads to
\beqar\label{cdiag16}
e^4 \de \FF{\si,\sixteen}{2}&=&
\coeffsymbol_{\si,\sixteen}^{ A Z}
\deDD{\sixteen}(\pmass,\MW,\MZ)
,
\eeqar
where 
\beqar\label{c2diag16}
\coeffsymbol_{\si,\sixteen}^{ A Z}
&=&
(D-1)
e^2 \gw^2
\left[
I_\si^{ A}
I_\si^{Z}
\sum_{a_2=\pm}
\left(
\de^{\SUtwo}_{ A Z}
\de^{\SUtwo}_{\bar{a}_2 a_2}
-
\de^{\SUtwo}_{ A a_2}
\de^{\SUtwo}_{\bar{a}_2 Z}
\right)
+(A\leftrightarrow Z)
\right]
\nl&=&
2(D-1)
(\dimtwo-1)
e^2 \gw^2 \sw\cw
Q I_\si^Z
,
\eeqar
and $\dimtwo=3$ is the dimension of the $\SUtwo$ group.

\subsubsection{Diagram \fif}
\beqar\label{diagram15}
\normprojector{2}_\si\quad
\vcenter{\hbox{
\diagXV
}}
\;\equaldiag
\begin{array}{l}
\\
\displaystyle{
e^6 \vev^2
 \FF{\si}{0}
\sum_{a_1,a_3,a_4
= A,Z,\pm} 
I_\si^{\bar{a}_1}
I_\si^{\bar{a}_4}
\sum_{\Phi_{i_2}=
H,\chi,\phi^\pm
}
\left\{
I^{{a}_1},
I^{\bar{a}_3}
\right\}_{H\Phi_{i_2}}
}
\\
\displaystyle{\quad{}\times
\left\{
I^{{a}_3},
I^{{a}_4}
\right\}_{\Phi_{i_2}H}
\DD{\fif}(M_{a_1},M_{\Phi_2},M_{a_3},M_{a_4})
,}
\end{array}
\nln
\eeqar
where 
the curly brackets denote anticommutators
and $\vev$ is the vev.
The loop integral $\DD{\fif}$ is defined in \refeq{defint15} and yields
\beqar\label{idiag15}
\MW^2\DD{\fif}(\MW,\MW,\MW,\MW)
&\NLLA&
\ResXVMMMM,
\nl
\MW^2\deDD{\fif}(M_1,M_2,M_3,M_4)
&\NLLA&
\ResdeXVMMMM,
\nl
\deDD{\fif}(\pmass,M_2,M_3,M_4)
&\NLLA&
\left(\frac{1}{{M_4}}\right)^2
\left[\ResdeXVZMMM\right],
\nl
\MW^2 \DD{\fif}(M_1,M_2,\pmass,M_4)
&\NLLA& 
\ResdeXVMMZM,
\nl
 \deDD{\fif}(M_1,M_2,M_3,\pmass)
&\NLLA&
\left(\frac{1}{{M_1}}\right)^2
\left[\ResdeXVMMMZ\right],
\nl
\MW^2
\DD{\fif}(\pmass,M_2,M_3,\pmass)
&\NLLA& 
\ResdeXVZMMZ.
\eeqar
This leads to
\beqar\label{cdiag15}
e^4 \de \FF{\si,\fif}{2}&=&
\coeffsymbol_{\si,\fif}^{ A Z}
\deDD{\fif}(\pmass,\MW,\MW,\MZ)
,
\eeqar
where 
\beqar\label{c2diag15}
\coeffsymbol_{\si,\fif}^{ A Z}
&=&
e^6 \vev^2
\left[
I_\si^{ A}
I_\si^{Z}
\sum_{a_3=\pm} 
\sum_{\Phi_{i_2}=
\phi^\pm
}
\left\{
I^{ A},
I^{\bar{a}_3}
\right\}_{H\Phi_{i_2}}
\left\{
I^{{a}_3},
I^{Z}
\right\}_{\Phi_{i_2}H}
+(A\leftrightarrow Z)
\right]
\nl&=&
-\frac{1}{2}e^2 \gw^4\frac{\sw}{\cw}
\vev^2
Q I_\si^Z
\left[\CA
-\left(\dimtwo-1\right)\cw^2
\right]
.
\eeqar
The above diagram is especially interesting 
since it represents the only contribution
involving  couplings  proportional to $\vev$,
which originate  from spontaneous symmetry
breaking.
One could naively expect that contributions involving 
such couplings vanish in the limit
$\vev^2/s\to 0$. However, in this case we see that the 
loop integrals associated to A--Z mixing-energy 
subdiagrams,
\beqar
\DD{\fif}(\pmass,\MW,\MW,\MZ)
&\NLLA&
\left(\frac{1}{{\MZ}}\right)^2
\left[\ResdeXVMMMZ\right],
\eeqar
are inversely proportional to $\MZ^2$. 
As a result, the above diagram provides a non-suppressed 
correction in the high-energy limit.

\subsubsection{Diagram \twe}
\beqar\label{diagram12}
\normprojector{2}_\si\quad
\vcenter{\hbox{
\diagXII
}}
\equaldiag
\begin{array}{l}
\\
\displaystyle{
\frac{1}{2} e^4 
 \FF{\si}{0}
\sum_{a_1,a_4
= A,Z,\pm} 
I_\si^{\bar{a}_1}
I_\si^{\bar{a}_4}
\sum_{\Phi_{i_2},\Phi_{i_3}
=H,\chi,\phi^\pm
}
I^{{a}_1}_{\Phi_{i_3}\Phi_{i_2}}
I^{{a}_4}_{\Phi_{i_2}\Phi_{i_3}}
}
\\
\displaystyle{\quad{}\times
\DD{\twe}(M_{a_1},M_{\Phi_{i_2}},M_{\Phi_{i_3}},M_{a_4})
,}
\end{array}
\nln\eeqar
where the loop integral $\DD{\twe}$ is defined in \refeq{defint12} and yields
\beqar\label{massidiag12}
\DD{\twe}(\MW,\MW,\MW,\MW)
&\NLLA& \ResXIIMMMM,
\nl
\deDD{\twe}(M_1,M_2,M_3,M_4)
&\NLLA& \ResdeXIIMMMM,
\nl
\deDD{\twe}(\pmass,M_2,M_3,M_4)
&\NLLA& 
\left(\frac{M_2^2+M_3^2}{{2M_4^2}}\right)
\left[\ResdeXIIZMMM\right],
\nl
\deDD{\twe}(M_1,M_2,M_3,\pmass)
&\NLLA& 
\left(\frac{M_2^2+M_3^2}{{2M_1^2}}\right)
\left[\ResdeXIIMMMZ\right],
\nl
\deDD{\twe}(\pmass,M_2,M_3,\pmass)
&\NLLA& \ResdeXIIZMMZ,
\eeqar
for massive scalars inside the vacuum polarization%
\footnote{
For 
massless scalars inside the vacuum polarization
we would obtain a result analogous to massless fermions
(see \refse{se:fermions}):
$
4\DD{\twe}(m_1,0,0,m_4)
\NLLA
\DD{\nin,0}(m_1,0,0,m_4)$
for masses $m_1,m_4$ either equal to $0$ or of  the order of $\MW$.
}.
This leads to
\newcommand{\Mphipm}{M_{\phi^\pm}}
\beqar\label{cdiag12}
e^4 \de \FF{\si,\twe}{2}
&=&
\coeffsymbol_{\si,\twe}^{\sew}
\DD{\twe}(\MW,\Mphipm,\Mphipm,\MW)
+
\coeffsymbol_{\si,\twe}^{ A Z}
\deDD{\twe}(\pmass,\Mphipm,\Mphipm,\MZ)
\nl&&{}
+
\coeffsymbol_{\si,\twe}^{ A  A}
\deDD{\twe}(\pmass,\Mphipm,\Mphipm,\pmass),
\eeqar
where  
\beqar\label{c2diag12}
\coeffsymbol_{\si,\twe}^{\sew}
&=& e^4 \sum_{a_1,a_4
= A,Z,\pm} 
I_\si^{\bar{a}_1}
I_\si^{\bar{a}_4}
\,\Tr_\Phi\left(
I^{{a}_1}
I^{{a}_4}
\right )
=\gb^4 
\frac{Y_\Phi^2}{2}
\left(\frac{Y_\si}{2}\right)^2
+\frac{1}{2}\gw^4
\ctwo
,\nl
\coeffsymbol_{\si,\twe}^{ A Z}
&=&
\frac{1}{2} e^4 
\left[
I_\si^{ A}
I_\si^{Z}
\,
\sum_{\Phi_{i_2},\Phi_{i_3}
=\phi^\pm
}
I^{ A}_{\Phi_{i_3}\Phi_{i_2}}
I^{Z}_{\Phi_{i_2}\Phi_{i_3}}
+(A\leftrightarrow Z)
\right]
=
-e^2\sw\cw 
Q I_\si^{Z}
\left(
\gb^2 Y_\Phi^2-\gw^2
\right)
,\nl
\coeffsymbol_{\si,\twe}^{ A  A}
&=&
e^4 Q^2
\,\Tr_\Phi\left(
I^{ A}I^{ A}
\right )
=\frac{1}{2}e^2 Q^2
\left[
\gb^2\cw^2Y_\Phi^2+\gw^2\sw^2 
\right]
.
\eeqar
The traces $\Tr_\Phi$
in the above identities can be read off from
\refeq{betacoeffexp1}.

\subsubsection{Diagram \seventeen}
\beqar\label{diagram17}
\normprojector{2}_\si\quad
\vcenter{\hbox{
\diagXVII
}}
\;\;\equaldiag
\begin{array}{l}
\\
\displaystyle{
\frac{1}{2} e^4  
 \FF{\si}{0}
\sum_{a_1,a_3
= A,Z,\pm} 
I_\si^{\bar{a}_1}
I_\si^{\bar{a}_3}
\sum_{\Phi_{i_2}
=H,\chi,\phi^\pm
}
\left\{
I^{{a}_1}
,I^{{a}_3}
\right\}_{\Phi_{i_2}\Phi_{i_2}}
}
\\
\displaystyle{\quad{}\times
\DD{\seventeen}(M_{a_1},M_{\Phi_{i_2}},M_{a_3})
,}
\end{array}
\nln
\eeqar
where $\DD{\seventeen}\equiv \DD{\sixteen}$.
This leads to
\beqar\label{cdiag17}
e^4 \de \FF{\si,\seventeen}{2}&=&
\coeffsymbol_{\si,\seventeen}^{ A Z}
\deDD{\sixteen}(\pmass,\Mphipm,\MZ)
,
\eeqar
where 
\beqar\label{c2diag17}
\coeffsymbol_{\si,\seventeen}^{ A Z}
&=&
\frac{1}{2} e^4 
\left[
I_\si^{ A}
I_\si^{Z}
\sum_{\Phi_{i_2}=\phi^\pm
}
\left\{
I^{ A}
,I^{Z}
\right\}_{\Phi_{i_2}\Phi_{i_2}}
+(A\leftrightarrow Z)
\right]
=
2\coeffsymbol_{\si,\twe}^{ A Z}
.
\eeqar
Also this diagram, which yields NLL contributions
only through A--Z mixing-energy subdiagrams,
gives rise to a correction proportional to
$(\MW/\MZ)^2$ via $\deDD{\sixteen}(\pmass,\Mphipm,\MZ)$.
This correction cancels 
the contribution proportional to 
$(\MW/\MZ)^2$ that originates 
from diagram \twe, \ie the term
$\coeffsymbol_{\si,\twe}^{ A Z}\deDD{\twe}$ in
\refeq{cdiag12}.

\subsubsection{Diagram \nin}
\label{se:fermions}
Finally, we consider the contributions
involving fermionic self-energy subdiagrams.
Here we just consider a generic fermionic doublet $\Psi$
with components  $\Psi_i=u,d$.
The sum over the three generations 
of leptons and quarks, as well as colour factors, 
are implicitly understood.
Assuming that all down-type fermions are massless ($m_d=0$)
and that the masses of up-type fermions are $m_u= 0$ or $\Mt$,
we have
\beqar\label{diagram9}
\normprojector{2}_\si\quad
\vcenter{\hbox{
\diagIX 
}}
\equaldiag
\begin{array}{l}
\phantom{\Biggl\{}
\\
\phantom{\biggl\{}
\\
\displaystyle{
\frac{1}{2}
e^4  \FF{\si}{0}
\sum_{a_1,a_4
= A,Z,\pm} 
I_\si^{\bar{a}_1}
I_\si^{\bar{a}_4}
{}
\sum_{\rho=\pm}
\Biggl\{
\sum_{\Psi_i,\Psi_j=u,d}
\left(I^{{a}_1}_\rho\right)_{\Psi_j \Psi_i}
}
\\
\displaystyle{
{}\times
\left(I^{{a}_4}_\rho\right)_{\Psi_i \Psi_j}
\DD{\nin,0}(M_{a_1},m_i,m_j,M_{a_4})
}
\\
\displaystyle{
{}-
\left(I^{{a}_1}_\rho\right)_{uu}
\left(I^{{a}_4}_{-\rho}\right)_{uu}
m_u^2 \DD{\nin,m}(M_{a_1},m_u,m_u,M_{a_4})
\Biggr\}
,}
\end{array}
\nln
\eeqar
where $\DD{\nin,m}\equiv -4 \DD{\fif}$
represents the contribution associated to the $m_u$-terms in the
numerator of the up-fermion propagators, whereas the 
integral $\DD{\nin,0}$, which is defined in \refeq{defint9},
accounts for the remaining contributions.
This latter integral  yields
\beqar\label{idiag90}
\DD{\nin,0}(\MW,\MW,\MW,\MW)&\NLLA& \ResIXMMMM,
\nl
\deDD{\nin,0}(M_1,m_2,m_3,M_4)&\NLLA& \ResdeIXMMMM,
\nl
\deDD{\nin,0}(\pmass,m_2,m_3,M_4)&\NLLA& 
\left(\frac{m_2^2+m_3^2}{{2M_4^2}}\right)
\left[\ResdeIXZMMM\right],
\nl
\deDD{\nin,0}(M_1,m_2,m_3,\pmass)&\NLLA& 
\left(\frac{m_2^2+m_3^2}{{2M_1^2}}\right)
\left[\ResdeIXMMMZ\right]
,\nl
\deDD{\nin,0}(\pmass,M_2,M_3,\pmass)&\NLLA& \ResdeIXZMMZ
,\nl
\deDD{\nin,0}(\pmass,0,0,\pmass)&\NLLA& \ResdeIXZZZZ,
\eeqar
for masses $M_i\sim \MW$ and 
$m_2,m_3=0$ or $\Mt$.
Using the fact that 
\beqar
\deDD{\nin,0}(M_1,m_2,m_3,\pmass)
&=&
\frac{m_2^2+m_3^2}{2}
\deDD{\nin,m}(M_1,m_2,m_3,\pmass),
\eeqar
and a similar relation with $\pmass\leftrightarrow M_1$, we can write
\beqar\label{cdiag9}
\lefteqn{
e^4 \de \FF{\si,\nin}{2}
=
\coeffsymbol_{\si,\nin}^{\sew}
\DD{\nin,0}(\MW,\MW,\MW,\MW)
+
\coeffsymbol_{\si,\nin}^{ A Z}
m_u^2
\deDD{\nin,m}(\pmass,m_u,m_u,\MZ)
}\quad&&\\&&{}
+
\coeffsymbol_{\si,\nin}^{ A  A,0}
\deDD{\nin,0}(\pmass,0,0,\pmass)
+
\coeffsymbol_{\si,\nin}^{ A  A,m}
\left[
\deDD{\nin,0}(\pmass,m_u,m_u,\pmass)
-\deDD{\nin,0}(\pmass,0,0,\pmass)
\right]
,\nn
\eeqar
where%
\footnote{
For the generators of neutral-current interactions,
$I^V=I^A,I^Z$, 
we write
$
\left(I^{V}_\rho\right)_{\Psi_i \Psi_j}
=
\de_{\Psi_i \Psi_j}
\left(I^{V}_\rho\right)_{\Psi_i}
$. A similar notation is used for $Q_\rho=Q$ and for 
the hypercharge $Y_\rho$.
} 
\beqar
\coeffsymbol_{\si,\nin}^{\sew}
&=&
\frac{1}{2}\,
e^4 \sum_{a_1,a_4
= A,Z,\pm} 
I_\si^{\bar{a}_1}
I_\si^{\bar{a}_4}
\sum_{\rho=\pm}
\,\Tr_\Psi\left(
I^{{a}_1}_\rho
I^{{a}_4}_\rho
\right )
\nl&=&
\frac{1}{4}
\left[\gb^4 
\left(\frac{Y_\si}{2}\right)^2
\sum_{\rho=\pm}
\frac{
\left({Y}_{\rho}\right)^2_d
+\left({Y}_{\rho}\right)^2_u
}{2}+ \gw^4 
\ctwo
\right]
,\nl
\coeffsymbol_{\si,\nin}^{ A  A,0}
&=&
 e^4 Q^2
\left(
Q_u^2+Q_d^2
\right)
,\qquad
\coeffsymbol_{\si,\nin}^{ A  A,m}
=
 e^4 Q^2
Q_u^2
,
\eeqar
and the trace $\Tr_\Psi$
can be read off from
\refeq{betacoeffexp1}.
The $m_u^2\deDD{\nin,m}$-term in \refeq{cdiag9}
originates from 
A--Z mixing-energy contributions
to the 
$\DD{\nin,0}$- 
and $m_u^2\DD{\nin,m}$-terms  in \refeq{diagram9},
which are proportional to 
$(m_u/\MZ)^2$. Their combination yields
\beqar\label{c2diag9}
\coeffsymbol_{\si,\nin}^{ A Z}
&=&
\frac{1}{2}
e^4 
I_\si^{A}
I_\si^{Z}
{}
\sum_{\rho=\pm}
\Biggl[
\left(I^{A}_\rho\right)_u
\left(I^{Z}_\rho\right)_u
-
\left(I^{A}_\rho\right)_{u}
\left(I^{Z}_{-\rho}\right)_{u}
\Biggr]
+(A\leftrightarrow Z)
=
0
.
\eeqar
This means
the NLL contributions
 proportional to 
$(m_u/\MZ)^2$ 
vanish owing to a cancellation between 
the momentum- and mass-terms in the fermionic propagators.
This is  analogous to the cancellation between the 
$(\MW/\MZ)^2$-terms from A--Z mixing-energies 
in diagrams  \twe~and \seventeen.

\subsection{Renormalization}
In this section we discuss the 
renormalization of the  two-loop form factors.
As we will see, to NLL accuracy
only one-loop  counterterms  contribute.
These counterterms  are evaluated in a generic 
renormalization scheme, where the weak mixing angle
and the electromagnetic coupling constant
are renormalized at the scales $\muR$ and  $\mue$, respectively.
The on-shell (OS) renormalization scheme%
\footnote{
The NLL finite parts of the 
one-loop counterterms within the  OS scheme
can be found in  \citere{Pozzorini:rs}.}
corresponds to  $\muR=\MW$ and $\mue\le\MW$, 
whereas the minimal-subtraction (MS) scheme is simply obtained by setting
$\muR=\mue=\muD$, where $\muD$ is the scale of dimensional
regularization.
The  one-loop mass and coupling-constant counterterms
are expressed in terms of the
$\beta$-function coefficients
$\betacoeff{1}^{(1)}$,
$\betacoeff{2}^{(1)}$,
$\betacoeff{e}^{(1)}$,
and $\betacoeff{\QED}^{(1)}$,
which are defined in \refapp{se:betafunction}.

\subsubsection{Mass renormalization}
Since the tree-level form factors do not depend 
on the weak-boson masses,
the only two-loop contributions resulting from
the renormalization of the bare masses 
$M_{0,a}$
of the weak bosons $a=\PZ,\PW$,
\beqar\label{Mpertserie1}
M_{0,a}
&=&
M_a\left[1+
\sum_{l=1}^\infty
\left(\frac{\alpha}{4\pi}\right)^l
\normfact^l
\de Z^{(l)}_{M_a}\right],
\eeqar
are 
\beqar\label{massren}
\de^M \FF{\si}{2}&=&
\sum_{a=\PZ,\PW}
\de Z^{(1)}_{M_a} 
\,
\frac{\partial
\,\de \FF{\si}{1}
}{\partial \log M_a}
.
\eeqar
To NLL accuracy, the one-loop mass counterterms read
\beqar\label{oneloopMCTs}
e^2 \de Z^{(1)}_{M_a} 
&\NLLA&
-\frac{1}{2}
\left\{
e^2 \betacoeff{aa}^{(1)}
-4 \left[
\gb^2\left(\frac{Y_\Phi}{2}\right)^2
+\gw^2 C_\Phi\right]
+\gw^2\frac{3\Mt^2}{2\MW^2}
\right\}
\frac{1}{\veps}
\left(\frac{-s}{\muR^2}\right)^\veps,
\eeqar
where 
$\betacoeff{ZZ}^{(1)}
=\sw^2 \betacoeff{1}^{(1)}
+\cw^2 \betacoeff{2}^{(1)}
$ and
$\betacoeff{WW}^{(1)}
=\betacoeff{2}^{(1)}
$.
In order to determine the degree of singularity of the
two-loop mass-counterterm contribution \refeq{massren},
let us consider the LL contributions to the one-loop form factors,
\beqar\label{oneloppLLmass}
\de \FF{\si}{1}\sim M_a^j \veps^{k} \log^{2+k} (M_a),
\eeqar
where $k\ge -2$, and  also mass-suppressed 
contributions are included by assuming $j\ge 0$. 
The terms resulting from a mass renormalization 
of \refeq{oneloppLLmass} are of order 
\beqar\label{twoloppLLmass}
\de^M \FF{\si}{2}\sim 
M_a^j \veps^{k} \left[
(2+k)\log^{2+k-1} (M_a)
+j \log^{2+k} (M_a)\right]
\de Z^{(1)}_{M_a}
\eeqar
and, as we see, only non-suppressed contributions with $j=0$ or 
$j=\order(\veps)$ remain non-suppressed.
Moreover, since the mass counterterms \refeq{oneloopMCTs}
provide only poles of order $\veps^{-1}$, we conclude that the 
degree of singularity of \refeq{twoloppLLmass} is only $2$ and
\beqar\label{nomassren}
\de^M \FF{\si}{2}&\NLLA&0
.
\eeqar

\subsubsection{Coupling-constant renormalization}
Since the tree-level form factors do not depend 
on the EW coupling constants,
the only two-loop contributions 
resulting from the 
renormalization
\beqar\label{PRpertserie1}
g_{0,i}
&=&
g_i+\sum_{l=1}^\infty
\left(\frac{\alpha}{4\pi}\right)^l
\normfact^l
\de g^{(l)}_i
,\qquad
e_0
=
e+\sum_{l=1}^\infty
\left(\frac{\alpha}{4\pi}\right)^l
\normfact^l
\de e^{(l)}
\eeqar
of the bare couplings  
$g_{0,1}$,
$g_{0,2}$,
and $e_{0}$, are
\beqar
e^2 \de^{\mathrm{CR}}\FF{\si}{2}&=&
\left(\sum_{i=1}^2
\de g_{i}^{(1)} 
\frac{\partial}{\partial g_i}
+
\de e^{(1)} 
\frac{\partial}{\partial e}
\right)
\left[
e^2\de \FF{\si}{1}
\right]
.
\eeqar
Within the on-shell scheme \cite{Denner:kt}, 
the renormalized parameters $e,\gb,\gw,\cw,\MZ,$ and $\MW$,
are fixed by renormalization conditions
for the  electromagnetic charge and the 
weak-boson masses and through  
the relations 
\refeq{massrelation},
\refeq{neutralhiggs},
and \refeq{genmixing}.
The counterterms for the renormalization 
of the bare 
weak mixing angle, 
\beqar\label{PRpertserie2}
\cwbare
&=&
\cw+
\sum_{l=1}^\infty
\left(\frac{\alpha}{4\pi}\right)^l
\normfact^l
\de \cw^{(l)}
,
\qquad
\swbare
=
\sw+\sum_{l=1}^\infty
\left(\frac{\alpha}{4\pi}\right)^l
\normfact^l
\de \sw^{(l)}
,
\eeqar
result from the weak-boson mass counterterms \refeq{oneloopMCTs}
and read
\beqar\label{cwcouplingren}
\de \cw^{(1)} &\NLLA&
\frac{\cw \sw^2}{2}
\left(
\betacoeff{1}^{(1)}
-\betacoeff{2}^{(1)}
\right)\frac{1}{\veps}
\left(\frac{-s}{\muR^2}\right)^\veps
,\qquad
\de \sw^{(1)} =
-\frac{\cw}{\sw}\,
\de \cw^{(1)}
.
\eeqar
The electric-charge renormalization at the scale 
$\mue\le\MW$ yields
\beqar\label{ecouplingren}
\de e^{(1)} &\NLLA&
-\frac{e}{2}
\frac{1}{\veps}
\left\{
\betacoeff{e}^{(1)}
\left(\frac{-s}{\muR^2}\right)^\veps
+
\betacoeff{\QED}^{(1)}
\left[
\left(\frac{-s}{\mue^2}\right)^\veps
-\left(\frac{-s}{\muR^2}\right)^\veps
\right]
\right\}
,
\eeqar
where the $\betacoeff{e}^{(1)}$-
and  $\betacoeff{\QED}^{(1)}$-terms
describe the running of the 
electromagnetic coupling 
above and below the electroweak scale,
respectively.
This latter is driven by 
light fermions only.  
The resulting counterterms for the gauge-couplings 
$\gb$ and $\gw$, read
\beqar\label{couplingren}
\de g_{i}^{(1)} &\NLLA&
-\frac{g_i}{2}
\frac{1}{\veps}
\left\{
\betacoeff{i}^{(1)}
\left(\frac{-s}{\muR^2}\right)^\veps
+
\betacoeff{\QED}^{(1)}
\left[
\left(\frac{-s}{\mue^2}\right)^\veps
-\left(\frac{-s}{\muR^2}\right)^\veps
\right]
\right\}
.
\eeqar

\subsubsection{Wave-function renormalization}
The fermionic WFRC's \refeq{WFpertserie1} 
yield the following two-loop counterterm-contributions 
\beqar\label{WFren2a}
\de^{\mathrm{WF}} 
\FF{\si}{2}&=&
\de Z^{(1)}_{\si} \de \FF{\si}{1}+
\de Z^{(2)}_{\si} 
-\left(\de Z^{(1)}_{\si}\right)^2 
,
\eeqar
where $\de \FF{\si}{1}$ are the renormalized one-loop form factors.
Since both $\de Z^{(2)}_\si$ 
and $\left(\de Z^{(1)}_\si\right)^2$ 
provide only poles of order $\veps^{-2}$,
to NLL accuracy we simply have
\beqar\label{WFren2b}
\de^{\mathrm{WF}} 
\FF{\si}{2}&\NLLA&
\de Z^{(1)}_{\si} \de \FF{\si}{1},
\eeqar
and the one-loop WFRC's are given in \refeq{WFRC}.

\subsection{Renormalized two-loop form factors}
In this section, we present the
results for the
renormalized two-loop form factors
\beqar\label{twoloopcomb}
\de \FF{\si}{2}&\NLLA&\sum_{i=1}^{11}
\de \FF{\si,i}{2}+
\de^{\mathrm{WF}}\FF{\si}{2}
+
\de^{\mathrm{CR}}\FF{\si}{2}.
\eeqar
Various aspects of these
results are discussed in detail in
\refse{se:discussion}.

\subsubsection{Analogy with Catani's formula}
As we show below,
the  LL and NLL corrections \refeq{twoloopcomb}
can be written in a form that is 
analogous to 
Catani's formula 
for two-loop mass singularities in massless QCD \cite{Catani:1998bh}.
In particular, we find that the two-loop corrections can be expressed in
terms of the one-loop functions 
$\univfact{\veps}{m}$ and
$\Delta\univfact{\veps}{m}$ defined in 
\refeq{Ifunc}--\refeq{subtractionofI}.
In the massless case ($m=0$), these functions are related to the factors 
${\bf I}^{(1)}$ defined in Eq.~(13) of \citere{Catani:1998bh}.
Actually, for 
a matrix element involving an external singlet gauge boson
and two charged fermions,
in the abelian case
($C_F=Q^2$,
$\CA=0$,
$T_R=1$)
\beqar\label{catanicomp2}
\frac{\alpha_S(\mu^2)}{2\pi}\,
{\bf I}^{(1)}(k \veps,\mu^2;\{p\})
\frac{\Gamma(1-k \veps)}{\Gamma(1-\veps)}\,
e^{(1-k)\gamma_{\mathrm{E}}\veps}
&=&
\frac{\alpha}{4\pi}
\normfact
Q^2\univfact{k\veps}{0}
\left(\frac{-s}{\mu^2}\right)^{(1-k)\veps}
,
\eeqar
at the renormalization scales $\mu=\mue=\muR$, where
the coupling constant 
of  \citere{Catani:1998bh}
is related to $\alpha$
by
\beqar\label{catanicomp1}
\alpha_S(\mu^2)
&=&
\left(\frac{4\pi\mu_0^2}{\mu^2}\right)^\veps
e^{-\gamma_{\mathrm{E}}\veps}
\,
\alpha
,
\eeqar
to one-loop accuracy, 
whereas for the corresponding 
$\beta$-function coefficients we have 
$4\pi \beta_0=
\betacoeff{e}^{(1)}
$.
In order to write the two-loop LL and NLL corrections
in a compact way,
it is useful to define
\beqar\label{NLLfacts}
\NLLfact{\veps}{m}{\mu}&:=&
\frac{1}{\veps}\left[
\univfact{2\veps}{m}
-\left(\frac{-s}{\mu^2}\right)^\veps
\univfact{\veps}{m}
\right]
\eeqar
and the corresponding subtracted functions
\beqar\label{subNLLfacts}
\Delta \NLLfact{\veps}{m}{\mu}&:=&
\NLLfact{\veps}{m}{\mu}
-
\NLLfact{\veps}{\MW}{\mu}.
\eeqar
In the massless abelian case,
the function \refeq{NLLfacts}
is related to the combination of 
 ${\bf I}^{(1)}$ functions 
that corresponds to the 
two-loop next-to-leading mass singularities proportional to 
$\beta_0$ in 
Eq.~(19) of \citere{Catani:1998bh},
\beqar\label{catanicomp3}
&&\left[\frac{\alpha_S(\mu^2)}{2\pi}\right]^2\,
\frac{2\pi\beta_0}{\veps}
\left[
{\bf I}^{(1)}(2 \veps,\mu^2;\{p\})
\frac{\Gamma(1-2 \veps)}{\Gamma(1-\veps)}\,
e^{-\gamma_{\mathrm{E}}\veps}
-
{\bf I}^{(1)}(\veps,\mu^2;\{p\})
\right]
=
\nl&&\quad =
\left[\frac{\alpha}{4\pi}\normfact\right]^2
Q^2
\betacoeff{e}^{(1)}
\NLLfact{\veps}{0}{\mu}
\Gamma(1-\veps)e^{-\gamma_{\mathrm{E}}\veps}
\NLLA
\left[\frac{\alpha}{4\pi}\normfact\right]^2
Q^2
\betacoeff{e}^{(1)}
\NLLfact{\veps}{0}{\mu}
.
\eeqar

\subsubsection{Results}
\label{se:finalewres}
We find that,
at the renormalization scales  $\mue=\muR=\MW$,
the two-loop form factors \refeq{twoloopcomb} can be written 
in terms of the one-loop form factors \refeq{onelooprena}--\refeq{subtractionofI},
\beqar\label{oneloopren}
e^2 \de \FF{\si}{1}&\NLLA&
\left[
\gb^2\left(\frac{Y_\si}{2}\right)^2
+\gw^2\ctwo
\right]
\univfact{\veps}{\MW}
\nl&&{}+
\left[
\gb^2\left(\frac{Y_\si}{2}\right)^2 
+\gw^2(T_\si^3)^2-e^2 Q^2
\right]
\Delta
\univfact{\veps}{\MZ}
+
e^2 Q^2
\Delta\univfact{\veps}{0},
\eeqar
and the functions \refeq{NLLfacts}--\refeq{subNLLfacts}
as 
\beqar\label{ewresummed1}
\left.
e^4 \de \FF{\si}{2}\right|_{\mu_i=\MW}
&\NLLA&
\frac{1}{2}
\left[e^2 \de \FF{\si}{1}
\right]^2
+e^2 \left[
\gb^2\betacoeff{1}^{(1)}
\left(\frac{Y_\si}{2}\right)^2
+\gw^2\betacoeff{2}^{(1)}
\ctwo
\right]
\NLLfact{\veps}{\MW}{\MW}
\nl&&{}
+
e^4 \betacoeff{\QED}^{(1)}
Q^2
\Delta \NLLfact{\veps}{0}{\MW}
.
\eeqar
Here, the first term on the right-hand side
can be regarded as the result of
the exponentiation of the  one-loop form factors.
This term contains the products of 
one-loop functions
\beqar\label{explewresNLL}
\left[
\univfact{\veps}{\MW}
\right]^2&\NLLA&
L^4-6L^3
+\order(\veps)
,\nl
\left[
\Delta\univfact{\veps}{\MZ}
\right]^2&\NLLA&
0
,\nl
\left[\Delta\univfact{\veps}{0}\right]^2
&\NLLA&
4\veps^{-4}
-4L^2\veps^{-2}
-\frac{8}{3}L^3\veps^{-1}
+12\veps^{-3}
+12L\veps^{-2}
-8L^3
+\order(\veps)
,\nl
\univfact{\veps}{\MW}
\Delta\univfact{\veps}{0}
&\NLLA&
2L^2\veps^{-2}
+\frac{4}{3}L^3\veps^{-1}
-\frac{1}{2}L^4
-6L\veps^{-2}
+7L^3
+\order(\veps)
,\nl
\univfact{\veps}{\MW}
\Delta\univfact{\veps}{\MZ}
&\NLLA&
-2
\LMZW L^3
+\order(\veps)
,\nl
\Delta\univfact{\veps}{\MZ}
\Delta\univfact{\veps}{0}
&\NLLA&
-4\LMZW\left[
L\veps^{-2}
+L^2\veps^{-1}
\right]
+\order(\veps)
,
\eeqar
whereas the 
functions $J$ and  $\Delta J$,
which are associated to the one-loop 
$\beta$-function coefficients,
 yield the NLL contributions
\beqar\label{explewresNLLb}
\NLLfact{\veps}{\MW}{\MW}
&\NLLA&
\frac{1}{3}L^3
+\order(\veps)
,\nl
\Delta \NLLfact{\veps}{0}{\MW}
&\NLLA&
\frac{3}{2}\veps^{-3}
+{2}L\veps^{-2}
+L^2\veps^{-1}
+\order(\veps)
.
\eeqar
The dependence 
of the two-loop form factors 
on the  renormalization scales 
$\mue$ and $\muR$, which 
originates from the 
counterterms \refeq{ecouplingren}--\refeq{couplingren},
is easily obtained by shifting
the couplings in the one-loop form factors.
This results into 
\beqar\label{rsdepresummed1}
\lefteqn{
e^4 \de \FF{\si}{2}-
\left[
e^4 \de \FF{\si}{2}\right]_{\mu_i=\MW}
\NLLA
}\quad&&\nl&\NLLA&
e^2\left[
\gb^2\betacoeff{1}^{(1)}
\left(\frac{Y_\si}{2}\right)^2
+\gw^2\betacoeff{2}^{(1)}
\ctwo
\right]
\left[
\NLLfact{\veps}{\MW}{\muR}
-
\NLLfact{\veps}{\MW}{\MW}
\right]
\nl&&{}+
e^4\betacoeff{e}^{(1)}  
Q^2
\left[
\Delta \NLLfact{\veps}{0}{\muR}
-
\Delta\NLLfact{\veps}{0}{\MW}
\right]
\nl&&{}+
e^2\betacoeff{\QED}^{(1)} 
\Biggl\{
\left[
\gb^2\left(\frac{Y_\si}{2}\right)^2
+\gw^2\ctwo
\right]
\left[
\NLLfact{\veps}{\MW}{\mue}
-
\NLLfact{\veps}{\MW}{\muR}
\right]
\nl&&{}+
e^2 
Q^2
\left[
\Delta \NLLfact{\veps}{0}{\mue}
-
\Delta\NLLfact{\veps}{0}{\muR}
\right]
\Biggr\}
.
\eeqar
Within the OS scheme, where  $\muR=\MW$, 
\refeq{rsdepresummed1}
reduces to the last
two lines with 
\beqar\label{ZgapI2NLLb}
\NLLfact{\veps}{\MW}{\mue}
-
\NLLfact{\veps}{\MW}{\MW}
&\NLLA&
-\Lmue L^2
+\order(\veps)
,\\
\Delta\NLLfact{\veps}{0}{\mue}
-
\Delta\NLLfact{\veps}{0}{\MW}
&\NLLA&
-\Lmue 
\left[
2\veps^{-2}
+\left(2L-\Lmue\right)\veps^{-1}
-\Lmue\left(L-\frac{1}{3}\Lmue\right)
\right]
+\order(\veps)
.\nn
\eeqar

\section{Discussion}
\label{se:discussion}
In this section, we discuss various features of the results
\refeq{ewresummed1}--\refeq{ZgapI2NLLb} and compare them
with \citeres{Fadin:2000bq,Kuhn:2000nn,Melles:2001gw,Catani:1998bh}.
In particular, we focus on the effects originating 
from various aspects of spontaneous symmetry breaking,
such as gauge-boson mixing, 
mass gaps and couplings with mass dimension.
To this end, 
we first consider various subsets of
the EW two-loop corrections 
corresponding to simpler gauge theories,
such as massless
 QED or an $\SUtwo\times\Uone$ theory with gauge-boson masses 
$M_A=\MZ=\MW$,
where these aspects are absent or only partially present.

\label{se:twoloodisc}
\subsection{Special cases}
Our final result,
as well as the contributions from individual Feynman diagrams and
counterterms in \refses{se:oneloop} and \ref{se:twoloop},
and \refapp{se:betafunction},
have been written in such a form that 
the following  special cases can 
be obtained by 
simple substitutions.
\newcommand{\subhiggsless}{(i)}
\newcommand{\sublighttop}{(ii)}
\newcommand{\subunmixed}{(iii)}
\newcommand{\subsimplegroup}{(iv)}
\newcommand{\subabelian}{(v)}
\newcommand{\subnowzgap}{(vi)}
\newcommand{\subheabyphoton}{(vii)}
\begin{enumerate}
\item[\subhiggsless]
We can consider a Higgsless theory by 
omitting the diagrams \fif--\seventeen, as well as the scalar contributions to the $\beta$-function, \ie
\beqar\label{sub:higgsless} 
\betacoeff{\Phi,1}^{(1)}&=&
\betacoeff{\Phi,2}^{(1)}=
\betacoeff{\Phi,e}^{(1)}=0
.
\eeqar

\item[\sublighttop]
We can treat the top quark as massless by 
setting
\beqar
\Mt=0,
\qquad
\betacoeff{\QED}^{(1)}=\betacoeff{\mathrm{F},e}^{(1)},
\eeqar
so that in contrast to 
\refeq{betacoeff2}, which 
corresponds to $\Mt\sim\MW$, 
also the 
top quark  contributes
to the  QED $\beta$-function.

\item[\subunmixed] 
As discussed in \refse{se:gaugeint}, the mixing effects can be consistently
removed by setting 
\beqar\label{sub:unmixed} 
Y_\Phi&=&0
,\qquad
\sw=0
,\qquad
\cw=1
,\qquad
\MZ=\MW.
\eeqar
In this case, the photon decouples from the 
weak bosons as well as from the Higgs doublet, since 
all components of $\Phi$ 
have electric charge $eQ_\Phi=\gb Y_\Phi/2=0$.
As a consequence, the 
photonic contributions to the diagrams \thr~and 
\ten--\seventeen~vanish. 
In particular, the diagrams  \sixteen, \fif, and \seventeen,
which provide NLL contributions only via  A--Z mixing-energies,
do not contribute at
all.

\item[\subsimplegroup] 
After the mixing has been removed, one can restrict oneself 
to a simple $\Uone$  gauge group with a massless photon 
by setting
\beqar
\gw=0
,\qquad
\gb=e
,\qquad
Y/2=Q
,\qquad
\betacoeff{1}^{(1)}=
\betacoeff{e}^{(1)}
.
\eeqar

\item[\subabelian] 
The nonabelian $\SUtwo\times\Uone$ 
interactions can be replaced by abelian $\Uone\times\Uone$ interactions,
with a massless photon and a massive Z boson,
by setting
\beqar\label{sub:abelian} 
C_\si&=&(T^3_\si)^2
,\qquad
\CA=0
,\qquad
\dimtwo=1
,\qquad
L=\log\left(\frac{-s}{\MZ^2}\right)
,\qquad
\LMZW=0.
\eeqar
As a consequence, 
the diagrams \thr~and \ten--\fif, as well as  
the 
gauge-boson and ghost contributions to the $\beta$-function, 
$\betacoeff{V,1}^{(1)}$,
$\betacoeff{V,2}^{(1)}$,
and
$\betacoeff{V,e}^{(1)}$,
vanish.

\item[\subnowzgap] 
The effects originating from the gap between the 
Z- and the W-boson mass in the weak-boson propagators%
\footnote{
In order to distinguish the 
weak-boson masses occurring in the propagators 
from those occurring in couplings with mass dimension, these latter
have been written in terms of $\vev$.
}
can be removed by setting
\beqar\label{sub:nogapWZ} 
\MZ=\MW
\eeqar
in all diagrams as well as in the WFRC's. This implies
$\LMZW=0$. 
When we study the effects of \refeq{sub:nogapWZ} 
on the diagrammatic calculation
we do not use
the identity 
\refeq{massrelation}, which is
violated in 
the case of  a non-vanishing 
mixing angle $(\cw\neq 1)$.

\item[\subheabyphoton]
Finally, one can consider the case where one sets the same mass $\MW$
in the propagators of all gauge bosons, A, Z, and $\PWpm$.
As discussed in \refse{se:massgapeff}, 
this is obtained by the substitution
\beqar\label{sub:nogap} 
Q&=&0,
\eeqar
combined with \refeq{sub:nogapWZ}.
In this case, the diagrams \sixteen, \fif, and \seventeen~vanish.

\end{enumerate}
In the following, we apply various combinations of the above substitutions
to each individual Feynman diagram 
and discuss their effect on the resulting two-loop corrections \refeq{twoloopcomb}.

\subsubsection{Massless QED}
\label{se:masslessQED}
The case of massless QED is obtained 
by combining the substitutions
\subhiggsless--\subsimplegroup. 
In this case, the one-loop form factors read
\beqar\label{oneloopren1}
e^2 \de \FF{\si}{1}&\NLLA&
e^2 Q^2
\univfact{\veps}{0}
,
\eeqar
and at two loops we obtain
\beqar\label{resummed1}
e^4 \de \FF{\si}{2}&\NLLA&
\frac{1}{2}
e^4 Q^4\left[
\univfact{\veps}{0}
\right]^2
+
e^4 Q^2
\betacoeff{\QED}^{(1)}
\NLLfact{\veps}{0}{\mue}
,
\eeqar
where 
$\betacoeff{\QED}^{(1)}
=\betacoeff{e}^{(1)}
=\betacoeff{\mathrm{F},e}^{(1)}$,
and to NLL accuracy  
\beqar\label{MasslessI2NLL}
\left[
\univfact{\veps}{0}
\right]^2&\NLLA&
4\Epsinv{4}
+12\Epsinv{3}
,\nl
\NLLfact{\veps}{0}{\mu_i}
&\NLLA&
\frac{3}{2}\Epsinv{3}
+{2}(L-\Lmui)^{}\Epsinv{2}
+(L-\Lmui)^{2}\Epsinv{1}
+\frac{1}{3}(L-\Lmui)^{3}
+\order(\veps)
.\eeqar
By means of 
\refeq{catanicomp2}
and \refeq{catanicomp3},
it is easy to see that the above  result 
agrees with 
the abelian case of
Catani's formula  for massless QCD \cite{Catani:1998bh}.

\subsubsection{Massive $\SUtwo\times\Uone$ theory}
\label{se:massivesymm}
Let us consider the case where 
we suppress the gauge-boson mass gaps
by replacing all gauge-boson masses with  $\MW$
in the propagators, 
but keeping  all other features of the EW theory
unchanged.
In this case,
which is obtained by  the substitutions 
\subnowzgap~and \subheabyphoton, 
the one-loop form factors read 
\beqar\label{oneloopren5}
e^2 \de \FF{\si}{1}&\NLLA&
\left[\gb^2\left(\frac{Y_\si}{2}\right)^2
+\gw^2\ctwo
\right]
\univfact{\veps}{\MW}
,
\eeqar
and at two loops,
for $\mue=\muR=\MW$, we obtain
\beqar\label{massivesymmresummed}
\left.
e^4 \de \FF{\si}{2}\right|_{\mu_i=\MW}
&\NLLA&
\frac{1}{2}
\left[\gb^2\left(\frac{Y_\si}{2}\right)^2
+\gw^2\ctwo
\right]^2
\left[\univfact{\veps}{\MW}\right]^2
\nl&&{}
+e^2\left[
\gb^2\betacoeff{1}^{(1)}\left(\frac{Y_\si}{2}\right)^2
+\gw^2\betacoeff{2}^{(1)}\ctwo
\right]
\NLLfact{\veps}{\MW}{\MW}
.
\eeqar
We note that
the two-loop NLL contributions that are 
proportional to the $\beta$-function coefficients
in \refeq{massivesymmresummed}
can be written,
analogously to the massless case \refeq{resummed1},
 in terms of the function $J(\veps,\MW,\MW)$,
which involves a factor $\veps^{-1}$ in its definition
\refeq{NLLfacts}.
This indicates that the two-loop NLL's of the type 
$\betacoeff{i}^{(1)}L^3$ 
are related to the 
one-loop LL's of order $\veps L^3$ in $\univfact{\veps}{\MW}$.

We also note that the above  result 
is independent of the gauge-boson mixing as well as of the 
symmetry-breaking mechanism.
In fact, if we compute all diagrams in the unmixed 
case \subunmixed,  if we completely remove the Higgs
doublet
with the substitutions \subhiggsless, or if we simply neglect
symmetry breaking ($\vev=0$), the resulting two-loop form factors 
remain unchanged.
In particular,
in absence of the mass gaps between the gauge bosons, 
each of the diagrams \one--\nin~is 
independent of the weak mixing angle.
Moreover, 
the only diagram that is proportional to the vev, 
\ie diagram \fif,
does not contribute.

Using \refeq{explewresNLL}--\refeq{explewresNLLb}
we see that, 
in the special cases of pure SU(2) and pure U(1) massive theories,
which result from $\gb=0$ and $\gw=0$, respectively,
the result \refeq{massivesymmresummed}
is in agreement with \citere{Kuhn:2000nn}.

\subsubsection{Massless U(1) theory with heavy top quark and scalar
doublet}
\label{se:heavyqed}
In order to discuss the effect resulting from 
the gap between the (vanishing) mass of the photon 
and the heavy 
masses of the top quark and scalar particles,  
let us consider the  case 
of a $\Uone$ theory
with fermions and a scalar doublet,
which is obtained by  removing 
symmetry breaking ($\vev=0$) and combining the substitutions%
\footnote{
In this case,  we
do not use \refeq{neutralhiggs}
and keep 
$Y_\Phi\neq 0$ in \subunmixed,
so that the components of the 
scalar doublet have a non-vanishing  
electric charge, 
$eQ_\Phi=\gb Y_\Phi/2$.
}
\subunmixed~and \subsimplegroup.
In this case, the one-loop form factors are as in 
\refeq{oneloopren1}
and at two loops, for $\mue=\muR=\MW$, we obtain
\beqar\label{resummed1a}
\left.
e^4 \de \FF{\si}{2}
\right|_{\mu_i=\MW}
\NLLA
\frac{1}{2}
e^4 Q^4\left[
\univfact{\veps}{0}
\right]^2
+
e^4 
Q^2
\left[
\betacoeff{e}^{(1)}
\NLLfact{\veps}{\MW}{\MW}
+
\betacoeff{\QED}^{(1)}
\Delta\NLLfact{\veps}{0}{\MW}
\right]
.\nln
\eeqar
As one can easily see by using the definition \refeq{subNLLfacts},
the only difference 
between \refeq{resummed1a} and 
the result \refeq{resummed1}
for  massless QED consists of NLL's
proportional to 
$
\left(
\betacoeff{e}^{(1)}
-\betacoeff{\QED}^{(1)}
\right)$
times 
$\NLLfact{\veps}{\MW}{\MW}$.
Such terms, 
which result from diagrams \twe~and \seventeen,
from the top-quark contribution to diagram \nin,
and from the corresponding heavy-particle contributions to the $\beta$-function, 
do not give rise to infrared $1/\veps$ poles
and behave 
as if the photon would be
heavy [see \refeq{massivesymmresummed}].

\subsubsection{Mass-gap effects within an abelian  $\Uone\times\Uone$ theory}
\label{se:abeliangap}
In order to discuss the effects resulting from the gap 
between the vanishing mass of 
the photon ($\la=0$) and the weak-boson mass scale $\MW$,
let us consider a
$\Uone\times\Uone$ theory
with a massless photon and a Z boson with mass%
\footnote{Despite 
of the fact that there are no
W bosons in the abelian case, 
here we assume $\MZ=\MW$
since we have chosen $\MW$ as scale 
of the logarithms \refeq{logsymbol}
originating from massive gauge bosons.
} $\MZ=\MW$
coupled to 
(massless and heavy) fermions and to the Higgs doublet.
In this case, 
which results from the substitutions \subabelian,
the one-loop form factors read 
\beqar\label{onelooprengap}
e^2 \de \FF{\si}{1}&\NLLA&
\left[
\gb^2\left(\frac{Y_\si}{2}\right)^2
+\gw^2(T_\si^3)^2
\right]
\univfact{\veps}{\MW}
+
e^2 Q^2
\Delta\univfact{\veps}{0}
,
\eeqar
and at two loops, for $\mue=\muR=\MW$, we obtain
\beqar\label{abelianresummed1}
\left.
e^4 \de \FF{\si}{2}\right|_{\mu_i=\MW}
&\NLLA&
\frac{1}{2}
\left\{
\left[
\gb^2\left(\frac{Y_\si}{2}\right)^2
+\gw^2(T_\si^3)^2
\right]
\univfact{\veps}{\MW}
+
e^2 Q^2
\Delta\univfact{\veps}{0}
\right\}^2
\nl&&{}
+e^2 \left[
\gb^2
\betacoeff{1}^{(1)}
\left(\frac{Y_\si}{2}\right)^2
+\gw^2\betacoeff{2}^{(1)}
(T_\si^3)^2
\right]
\NLLfact{\veps}{\MW}{\MW}
\nl&&{}
+
e^4 \betacoeff{\QED}^{(1)}
Q^2
\Delta \NLLfact{\veps}{0}{\MW}
.
\eeqar
The various products of one-loop functions
that appear in this expression for the 
two-loop corrections have been expanded 
in \refeq{explewresNLL} and \refeq{explewresNLLb},
including terms up to the order $\veps^0$.
It is interesting to observe 
that certain one-loop functions
on the right-hand side of \refeq{abelianresummed1}
need to be expanded beyond order $\veps^0$
in order to reproduce all divergent and finite 
parts of the result 
\refeq{twoloopcomb}
of the diagrammatic NLL calculation.
Indeed, 
in the products of
the type $\univfact{\veps}{\MW}\Delta\univfact{\veps}{0}$,
\ie combinations of massive and massless
one-loop contributions, 
the massive one-loop terms have to be expanded 
up to the order $\veps^2$, since  
the massless one-loop terms involve poles of order
$\veps^{-2}$.
This means, for instance,
that 
the two-loop corrections of order $L^4$
in \refeq{abelianresummed1}
receive contributions from the one-loop corrections of order $\veps^{2}L^4$.

As in \refse{se:massivesymm}, 
also in the $\Uone\times\Uone$ case we find that the
final result
does not change if one performs the diagrammatic calculation
by neglecting the gauge-boson mixing and/or the 
symmetry-breaking mechanism.
In particular, the diagram \fif~does not contribute in the abelian case.

A detailed comparison 
of the result \refeq{abelianresummed1}
 with \citeres{Fadin:2000bq,Kuhn:2000nn,Melles:2001gw}
is not possible,
since we have regulated mass singularities 
dimensionally and we did not include real 
electromagnetic corrections.
However, we observe that \refeq{abelianresummed1},
and in particular its  dependence on the gauge-boson 
mass gap, $\la\ll\MW$,
confirms the approach that has been adopted in 
\citeres{Fadin:2000bq,Kuhn:2000nn,Melles:2001gw}
to resum the EW corrections. 
Indeed, we see that the only difference between 
\refeq{abelianresummed1} and the corresponding result 
for $\la=\MW$, 
\ie the abelian case of \refeq{massivesymmresummed},
consists of terms proportional to
$Q^2\Delta\univfact{\veps}{0}$ and 
$Q^2\betacoeff{\QED}^{(1)}\Delta\NLLfact{\veps}{0}{\MW}$,
which correspond to QED corrections
(with $\la=0$) subtracted at the scale $\la=\MW$.
This means that the two-loop corrections 
in presence of a mass gap $\la\ll\MW$
result from the corrections within an unbroken 
and unmixed gauge theory with $\la=\MW$
with additional mass-gap effects that 
are 
of pure electromagnetic nature.

\subsubsection{Mixing, Z--W mass gap and symmetry
breaking within the electroweak theory}
\label{se:ewresult}
Finally, let us discuss the results 
\refeq{ewresummed1}--\refeq{ZgapI2NLLb},
which are obtained by considering all two-loop diagrams within 
the spontaneously broken EW theory.
We first observe that the only  two-loop  contributions 
depending on the gaps between the heavy-particle masses
\refeq{simmasses}
are the NLL's proportional to  $\LMZW=\log(\MZ/\MW)$ that 
arise from the squared 
one-loop corrections in \refeq{ewresummed1}
via the 
$\Delta\univfact{\veps}{\MZ}$
subtracted terms in \refeq{oneloopren}.
This means that such Z--W mass-gap contributions
exponentiate.

If we neglect these  NLL's proportional to
$\LMZW$, 
the result \refeq{ewresummed1}
is  analogous to \refeq{abelianresummed1}
and
confirms the approach adopted 
in \citeres{Fadin:2000bq,Kuhn:2000nn,Melles:2001gw}
to resum the EW corrections, 
as discussed in \refse{se:abeliangap} for the abelian case.
In particular,
the effects resulting from the 
gap between the (vanishing) photon mass and the weak-boson 
mass scale turn out to be  of QED nature
also in  the non-abelian case.

We note that, apart from the 
$\LMZW$ contributions,
the result \refeq{ewresummed1} seem to be insensitive to
the gauge-boson mixing, the \PZ-\PW~mass gap, and
the symmetry breaking.
Indeed,
if we compute all diagrams 
in absence of mixing, with $\MW=\MZ$ in the weak-boson propagators, 
and $\vev=0$,
we obtain the same two-loop form factors
\refeq{ewresummed1}
with $\LMZW=0$.
However, when we  perform the calculation 
within the mixed spontaneously broken EW theory,
we observe that the mixing effects
drop out as a result of subtle cancellations 
between different diagrams,
which take place only if the 
\PZ-\PW~mass gap and the vev are properly taken into account.

In order to understand these mixing cancellations,
let us consider the behaviour of the 
diagrams  involving A--Z mixing-energies, \ie diagrams of the type
\beqar\label{diagramAZ}
\vcenter{\hbox{
\diagAZ
}},
\eeqar
which we call A--Z mixing diagrams.
The combination of 
all A--Z mixing diagrams \ten--\nin~yields
the  infrared-divergent contribution%
\footnote{Here we consider the infrared-divergent
$\Delta$-contributions 
corresponding to the 
subtracted terms in \refeq{subtractiondef}.}, 
\beqar\label{AZbosmixdelta1}
\Delta\de\FF{\si,\mathrm{AZ}}{2}
&\NLLA&
4 e^2 \gw^2 \sw \cw Q I^Z_\si
K(\MW,\MZ,\vev)
\left[
\veps^{-3}
+L\veps^{-2}
-\frac{2}{3}L^3\right],
\eeqar
where
\beqar\label{AZbosmixdelta2}
K(\MW,\MZ,\vev)
&=&
\left(\frac{\MW}{\MZ}\right)^2
\left[4\CA-3(\dimtwo-1)\right]
+\left(\frac{\gw\vev}{2\cw \MZ}\right)^2
\left[\CA-(\dimtwo-1)\cw^2\right]
\nl&=&
2\left\{
\left(\frac{\MW}{\MZ}\right)^2
+\sw^2 \left(\frac{\gw\vev}{2\cw \MZ}\right)^2
\right\}.
\eeqar
We note that only the bosonic diagrams \ten--\fif~contribute to
\refeq{AZbosmixdelta1}, whereas the fermionic A--Z mixing diagrams
\nin, and the  combination of scalar A--Z mixing diagrams
\twe~and \seventeen~are irrelevant.
The A--Z mixing contributions \refeq{AZbosmixdelta1} 
arise only in presence of mixing ($\sw\neq 0$) and are of
non-abelian nature,  
since the coefficient \refeq{AZbosmixdelta2} vanishes
within an abelian $\Uone\times\Uone$ theory,
where $\CA=0$ and $\dimtwo=1$.
Moreover, they involve the 
only (NLL) contribution that depends on the vev, \ie diagram \fif.

We see that, 
in contrast to the contributions from all other
diagrams, \refeq{AZbosmixdelta1}--\refeq{AZbosmixdelta2}
depend  linearly on the 
ratios $(\MW/\MZ)^2$ or  $(\vev/\MZ)^2$. 
For instance,  the A--Z mixing diagrams  yield
infrared-divergent terms
proportional to $(\MW/\MZ)^2\veps^{-3}$
that
arise as combinations of UV singularities 
proportional to $\MW^2\veps^{-1}$
from tadpole mixing-energy subdiagrams,
together with soft-collinear singularities proportional to 
$\MZ^{-2}\veps^{-2}$
from the region where the photon momentum 
vanishes and the Z-boson propagator equals $-\ri g^{\mu\nu}/\MZ^2$.

In the final result, we find that the A--Z mixing contribution
\refeq{AZbosmixdelta1} cancels 
against a corresponding mixing 
contribution proportional 
to $
\sw Q I^Z_\si
\left[
\veps^{-3}
+L\veps^{-2}
-2/3L^3\right]$
from all other diagrams\footnote{
Such contribution corresponds to 
\refeq{AZbosmixdelta1} with the factor \refeq{AZbosmixdelta2}
replaced by $-2$.}.
In order to ensure this mixing cancellation,
it is crucial 
to take into account the relations \refeq{massspectrum}
between the weak-boson masses and the vev.
In particular, one has to compute the A--Z mixing diagrams  using
$\MW=\cw\MZ$, \ie different masses in the weak-boson propagators.

If one would neglect the vev, \ie diagram \fif, or if one would
compute all diagrams with $\MW=\MZ$, 
the mixing cancellations would be destroyed, and 
one would find deviations
of order $\veps^{-3}$ with respect to \refeq{ewresummed1}.
However, it is interesting to note that
if one 
simultaneously
omits the diagram \fif~($\vev=0$) and 
one uses  $\MW=\MZ$ in the propagators,
then   \refeq{AZbosmixdelta2} remains unchanged,
\ie  the mixing cancellations are preserved
and one obtains the same result as in 
\refeq{ewresummed1}.
This means that the diagram \fif~cancels
the Z--W mass-gap effects  
resulting from the 
gauge-boson- and ghost- A--Z mixing diagrams \ten~and \sixteen.

\section{Conclusions}
\label{se:conclusions}

We have considered the  
one- and two-loop virtual electroweak (EW) corrections
to the form factors for an 
$\SUtwo\times \Uone$ 
singlet gauge boson of invariant mass $s$
coupling to massless chiral fermions.
In the asymptotic region $s\gg\MW^2$, 
we have computed mass singularities in $D=4-2\veps$ dimensions
taking into account leading logarithms (LL's)
and next-to-leading logarithms (NLL's).
This approximation includes
combinations $\alpha^l\veps^{k}\log^{j+k}(s/\MW^2)$
of mass-singular logarithms 
and  $1/\veps$ poles, with $j=2l,2l-1$, and  
$-j\le k\le 4-2l$.
The heavy particle masses, 
$\MZ\sim \MW\sim \Mt\sim \MH$, have been assumed to be of the same
order but not equal. 

We found that the EW two-loop LL's and NLL's
result from the exponentiation of the corresponding one-loop 
contributions plus additional NLL's that are proportional 
to the one-loop $\beta$-function coefficients.
This result has been expressed in a form that 
corresponds to a generalization of 
Catani's two-loop formula for massless QCD.

If one neglects 
the NLL's proportional to $\log(\MZ/\MW)$,
which turn out to exponentiate, 
the result confirms the approach 
that has  been adopted in
\citeres{Fadin:2000bq,Kuhn:2000nn,Melles:2001gw} 
to resum the EW corrections.
Indeed, 
the two-loop EW LL's and NLL's 
can be expressed through the corresponding
corrections within an unbroken and unmixed
$\SUtwo\times\Uone$  theory where 
all EW gauge bosons have mass $\MW$
and additional effects resulting from the
gap between the 
vanishing photon mass, $\la=0$, and the weak-boson mass scale $\MW$.
Such photonic mass-gap effects
turn out to consist of pure QED corrections, 
with $\la=0$, subtracted at the scale $\la=\MW$.

Apart from the $\log(\MZ/\MW)$-contributions,
the effects 
from the gauge-boson mixing, the \PZ-\PW~mass gap, and
the symmetry breaking cancel in the final result.
However, we found that this simple behaviour of the two-loop 
form factors
is ensured by subtle cancellations 
between subleading mass singularities from 
different diagrams where, 
in contrast to the assumptions (i)--(iii)
discussed in \refse{se:intro},
the details of  
spontaneous symmetry breaking 
cannot be neglected. 
In particular, we have stressed that 

\begin{itemize}
\item[(i)] Couplings with mass dimension, \ie proportional to the 
vacuum expectation value $\vev$,
cannot be neglected in the massless limit $\MW^2/s\to 0$. 

\item[(ii)] 
Diagrams involving A--Z mixing-energies,
which give rise to contributions proportional to
$(\MW/\MZ)^2$ and 
$(\vev/\MZ)^2$, 
have to be computed 
using different masses, $\MW\neq \MZ$, 
in the weak-boson propagators.

\item[(iii)]
The mixing effects drop out as a result of 
cancellations 
that  take place only if the 
relations between the weak-boson masses, the weak mixing angle  and
the vacuum expectation value 
are properly taken into account. 

\end{itemize}

We note that the symmetry-breaking effects 
were restricted to a small 
subset of diagrams in this form-factor calculation,
since  the Higgs sector is coupled to the 
massless fermions only via
one-loop  insertions in the gauge-boson
self-energies and mixing-energies.
Therefore, it would be interesting to 
extend the study
of two-loop subleading mass singularities 
to  processes involving 
heavy external particles, which are directly coupled
to the Higgs sector.

\section*{Acknowledgments}
I would like to thank A.~Denner, J.~H.~K\"uhn and O.~Veretin for fruitful 
discussions. I am grateful to A.~Denner for carefully reading the manuscript.
\begin{appendix}

\section{Loop integrals}\label{app:loops}
\newcommand{\lmom}{l}
\newcommand{\slmom}{\lslash}
\newcommand{\mass}{m}
\newcommand{\linea}[1]{k_{#1}}
\newcommand{\lineb}[1]{q_{#1}}
\newcommand{\slinea}[1]{\ks_{#1}}
\newcommand{\slineb}[1]{\qs_{#1}}
\newcommand{\measure}[1]{\rd \tilde{\lmom}_{#1}}
\newcommand{\propagatorm}[2]{P(#1,#2)}
\newcommand{\propagator}[1]{P(#1)}

In this appendix, we list the explicit expressions for the 
Feynman integrals that contribute to the 
one- and two-loop diagrams discussed in 
\refses{se:oneloop} and \ref{se:twoloop}.
In order to keep our expressions as compact as possible we
define the momenta
\beqar\label{momentadef}
\linea{1}&=&\lmom_1+p_1
,\qquad
\linea{2}=\lmom_2+p_1
,\qquad
\linea{3}=\lmom_1+\lmom_2+p_1
,\nl
\lineb{1}&=&\lmom_1-p_2
,\qquad
\lineb{2}=\lmom_2-p_2
,\qquad
\lineb{3}=\lmom_1+\lmom_2-p_2
,\qquad
\lmom_{3}=-\lmom_1-\lmom_2
,
\eeqar
for massive and massless propagators we use the notation 
\beq\label{propagator}
\propagatorm{q}{m}:=q^2-m^2+\ri 0
,\qquad
\propagator{q}:=q^2+\ri 0,
\eeq
and for triple gauge-boson couplings we write
\beqar\label{YMvrtex}
\Gamma^{\mu_1\mu_2\mu_3}(
\lmom_1,\lmom_2,\lmom_3)
&:=&
g^{\mu_1\mu_2}(\lmom_1-\lmom_2)^{\mu_3}
+g^{\mu_2\mu_3}(\lmom_2-\lmom_3)^{\mu_1}
+g^{\mu_3\mu_1}(\lmom_3-\lmom_1)^{\mu_2}
.
\eeqar
The normalization factors occurring in
\refeq{pertserie1} and \refeq{normproject}
are absorbed into the integration measure
\beq\label{measure}
\measure{i}:=
\frac{(4\pi)^2}
{\normfact} \muD^{4-D}
\frac{\rd^D \lmom_i}{\left(2\pi\right)^D}
=
\frac{1}{\pi^2} \,\Gamma\left(\frac{D}{2}-1\right)
(-s\pi)^{2-D/2}
\rd^D \lmom_i
, 
\eeq
and for the projection  introduced
in \refeq{projectordef}
we use the shorthand
\newcommand{\traceproject}{\mathrm{Pr}}
\beqar\label{traceproject}
\traceproject\left(\Gamma^\nu\right)
&:=&
\frac{1}{2(2-D)s}\Tr\left(\gamma_\nu \ps_1 \Gamma^\nu\ps_2 \right)
.
\eeqar
With this notation we have
\beqar
\label{defint0}
\DD{0}(\mass_1)&:=&
-\ri
\int \measure{1}
\frac{
\traceproject\left(
\gamma^{\mu_1}
\slinea{1}
\gamma^\nu
\slineb{1}
\gamma_{\mu_1}
\right)
}{
\propagatorm{\lmom_1}{\mass_1}
\propagator{\linea{1}}
\propagator{\lineb{1}}
}
,\nl
\label{defint1}
\DD{\one}(\mass_1,\mass_2)&:=&
-
\int \measure{1}  \measure{2}
\frac{
\traceproject\left(
\gamma^{\mu_1}
\slinea{1}
\gamma^{\mu_2}
\slinea{3}
\gamma^\nu
\slineb{3}
\gamma_{\mu_2}
\slineb{1}
\gamma_{\mu_1}
\right)
}{
\propagatorm{\lmom_1}{\mass_1}
\propagatorm{\lmom_2}{\mass_2}
\propagator{\linea{1}}
\propagator{\linea{3}}
\propagator{\lineb{1}}
\propagator{\lineb{3}}
}
,\nl
\label{defint2}
\DD{\two}(\mass_1,\mass_2)&:=&
-
\int \measure{1}  \measure{2}
\frac{\traceproject\left(
\gamma^{\mu_1}
\slinea{1}
\gamma^{\mu_2}
\slinea{3}
\gamma^\nu
\slineb{3}
\gamma_{\mu_1}
\slineb{2}
\gamma_{\mu_2}
\right)
}{
\propagatorm{\lmom_1}{\mass_1}
\propagatorm{\lmom_2}{\mass_2}
\propagator{\linea{1}}
\propagator{\linea{3}}
\propagator{\lineb{2}}
\propagator{\lineb{3}}
}
,\nl
\label{defint3}
\DD{\thr}(\mass_1,\mass_2,\mass_3)&:=&
\int \measure{1}  \measure{2}
\frac{
\traceproject\left(
\gamma^{\mu_1}
\slinea{1}
\gamma^{\mu_2}
\slinea{3}
\gamma^\nu
\slineb{3}
\gamma^{\mu_3}
\right)
\Gamma_{\mu_1\mu_2\mu_3}(
\lmom_1,\lmom_2,\lmom_3)
}{
\propagatorm{\lmom_1}{\mass_1}
\propagatorm{\lmom_2}{\mass_2}
\propagatorm{\lmom_3}{\mass_3}
\propagator{\linea{1}}
\propagator{\linea{3}}
\propagator{\lineb{3}}
}
,\nl
\label{defint5}
\DD{\fiv}(\mass_1,\mass_2)&:=&
-
\int \measure{1}  \measure{2}
\frac{\traceproject\left(
\gamma^{\mu_1}
\slinea{1}
\gamma^{\mu_2}
\slinea{3}
\gamma_{\mu_2}
\slinea{1}
\gamma^\nu
\slineb{1}
\gamma_{\mu_1}
\right)
}{
\propagatorm{\lmom_1}{\mass_1}
\propagatorm{\lmom_2}{\mass_2}
\left[\propagator{\linea{1}}\right]^2
\propagator{\linea{3}}
\propagator{\lineb{1}}
}
,\nl
\label{defint7}
\DD{\sev}(\mass_1,\mass_2)
&:=&
-
\int \measure{1}  \measure{2}
\frac{
\traceproject\left(
\gamma^{\mu_2}
\slinea{2}
\gamma^{\mu_1}
\slinea{3}
\gamma_{\mu_2}
\slinea{1}
\gamma^\nu
\slineb{1}
\gamma_{\mu_1}
\right)
}{
\propagatorm{\lmom_1}{\mass_1}
\propagatorm{\lmom_2}{\mass_2}
\propagator{\linea{1}}
\propagator{\linea{3}}
\propagator{\linea{2}}
\propagator{\lineb{1}}
}
,\nl
\label{defint10}
\DD{\ten}(\mass_1,\mass_2,\mass_3,\mass_4)&:=&
\int \measure{1}  \measure{2}
\frac{\traceproject\left(
\gamma^{\mu_1}
\slinea{1}
\gamma^\nu
\slineb{1}
\gamma_{\mu_4}
\right)
}{
\propagatorm{\lmom_1}{\mass_1}
\propagatorm{\lmom_2}{\mass_2}
\propagatorm{\lmom_3}{\mass_3}
\propagatorm{\lmom_1}{\mass_4}
\propagator{\linea{1}}
\propagator{\lineb{1}}
}\times
\nl&&{}\times\left[
\Gamma_{\mu_1\mu_2\mu_3}
(\lmom_1,\lmom_2,\lmom_3)
\Gamma^{\mu_4\mu_2\mu_3}
(\lmom_1,\lmom_2,\lmom_3)
+2\lmom_{2\mu_1}\lmom_{3}^{\mu_4}
\right]
,\nl
\label{defint16}
\DD{\sixteen}(\mass_1,\mass_2,\mass_3)&:=&
\int \measure{1}  \measure{2}
\frac{\traceproject\left(
\gamma^{\mu_1}
\slinea{1}
\gamma^\nu
\slineb{1}
\gamma_{\mu_1}
\right)
}{
\propagatorm{\lmom_1}{\mass_1}
\propagatorm{\lmom_2}{\mass_2}
\propagatorm{\lmom_1}{\mass_3}
\propagator{\linea{1}}
\propagator{\lineb{1}}
}
,\nl
\label{defint15}
\DD{\fif}(\mass_1,\mass_2,\mass_3,\mass_4)&:=&
\int \measure{1}  \measure{2}
\frac{\traceproject\left(
\gamma^{\mu_1}
\slinea{1}
\gamma^\nu
\slineb{1}
\gamma_{\mu_1}
\right)
}{
\propagatorm{\lmom_1}{\mass_1}
\propagatorm{\lmom_2}{\mass_2}
\propagatorm{\lmom_3}{\mass_3}
\propagatorm{\lmom_1}{\mass_4}
\propagator{\linea{1}}
\propagator{\lineb{1}}
},\nl
\label{defint12}
\DD{\twe}(\mass_1,\mass_2,\mass_3,\mass_4)&:=&
-
\int \measure{1}  \measure{2}
\frac{
\traceproject\left(
\gamma^{\mu_1}
\slinea{1}
\gamma^\nu
\slineb{1}
\gamma_{\mu_4}
\right)
(\lmom_{2}-\lmom_{3})_{\mu_1}
(\lmom_{2}-\lmom_{3})^{\mu_4}
}{
\propagatorm{\lmom_1}{\mass_1}
\propagatorm{\lmom_2}{\mass_2}
\propagatorm{\lmom_3}{\mass_3}
\propagatorm{\lmom_1}{\mass_4}
\propagator{\linea{1}}
\propagator{\lineb{1}}
}
,\nl
\label{defint17}
\DD{\seventeen}(\mass_1,\mass_2,\mass_3)
&:=&
\DD{\sixteen}(\mass_1,\mass_2,\mass_3)
,\nl
\label{defint9}
\DD{\nin,0}(\mass_1,\mass_2,\mass_3,\mass_4)&:=&
-
\int \measure{1}  \measure{2}
\frac{\traceproject\left(
\gamma^{\mu_1}
\slinea{1}
\gamma^\nu
\slineb{1}
\gamma_{\mu_4}
\right)
\Tr\left(
\gamma_{\mu_1}
\slmom_2
\gamma^{\mu_4}
\slmom_3\right)
}{
\propagatorm{\lmom_1}{\mass_1}
\propagatorm{\lmom_2}{\mass_2}
\propagatorm{\lmom_3}{\mass_3}
\propagatorm{\lmom_1}{\mass_4}
\propagator{\linea{1}}
\propagator{\lineb{1}}
}
,\nl
\label{defint9m}
\DD{\nin,m}(\mass_1,\mass_2,\mass_3,\mass_4)
&:=&
-4 \DD{\fif}(\mass_1,\mass_2,\mass_3,\mass_4)
.\nln
\eeqar
In order to isolate the effects originating from the mass gaps
as discussed in \refse{se:massgapeff}, 
we introduce the subtracted loop integrals
\beqar
\Delta \DD{k}(m_1,\dots,m_n):=
\DD{k}(m_1,\dots,m_n)-
\left[\DD{k}(m_1,\dots,m_n)\right]_{m_i=\MW}.
\eeqar

\section{$\beta$--function coefficients}
\label{se:betafunction}
In this appendix we give relations 
and explicit expressions for the 
one-loop $\beta$-function coefficients
$\betacoeff{1}^{(1)}$,
$\betacoeff{2}^{(1)}$,
$\betacoeff{e}^{(1)}$,
and $\betacoeff{\QED}^{(1)}$,
which have been used in the calculation.
For more details we refer to  \citere{Pozzorini:rs}.

The coefficients $\betacoeff{1}^{(1)}$, 
$\betacoeff{2}^{(1)}$  and 
$\betacoeff{e}^{(1)}$,
are related to the matrix 
\beqar\label{betamatrix1}
\betacoeff{ab}^{(1)}
=\betacoeff{1}^{(1)}
\de^{\groupa}_{{a} b}
+\betacoeff{2}^{(1)}
\de^{\groupb}_{{a} b}
,\qquad
a,b=A,Z,\pm,
\eeqar
which corresponds to the residues of  
ultraviolet poles of the 
gauge-boson self-energies and mixing-energies.
Here,
$\de^{\groupa}$ and $\de^{\groupb}$
are the projectors defined in \refeq{adjKron2}.
The coefficient 
corresponding to the electric-charge renormalization
is given by 
\beqar\label{betamatrixel}
\betacoeff{e}^{(1)}
=
\betacoeff{AA}^{(1)}
=\cw^2\betacoeff{1}^{(1)}
+\sw^2
\betacoeff{2}^{(1)}
.
\eeqar
Each $\beta$-function coefficient $
\betacoeff{k}^{(1)}
=\betacoeff{1}^{(1)}
,\betacoeff{2}^{(1)}
,\betacoeff{e}^{(1)}
,\betacoeff{ab}^{(1)}
$
receives contributions from the gauge ($V$), scalar ($\Phi$), and
fermionic (F) sectors,
and the fermionic contributions 
result from the sum over $N_G=3$ generations of leptons (L) and quarks (Q), 
with $N_c=3$ colours:
\beqar\label{betamatdef}
\betacoeff{k}^{(1)}
&=&
\sum_{\Rep=V,\Phi,\mathrm{F}}
\betacoeff{\Rep,k}^{(1)},
\qquad
\betacoeff{\mathrm{F},k}^{(1)}
=
N_G
\left[\betacoeff{\mathrm{L},k}^{(1)}
+N_c\,\betacoeff{\mathrm{Q},k}^{(1)}
\right].
\eeqar
The individual contributions to $\betacoeff{ab}^{(1)}$
read%
\footnote{
Each contribution 
$\betacoeff{\Rep,ab}^{(1)}$
is expressed in terms of the trace
\beqar\label{defDynkin}
\label{traceids}
e^2 \,\Tr_\Rep\left(
I^{\bar{a}}
I^{{b}}
\right )
=
e^2 
\sum_{\varphi_i,\varphi_j\in R}
I^{\bar a}_{\varphi_i\varphi_j}
I^b_{\varphi_j\varphi_i}
=\frac{1}{4}
\gb^2 
\Tr_\Rep\left(Y^2\right)
\de^{\groupa}_{{a} b}
+\gw^2 T_{\Rep}
\de^{\groupb}_{{a} b},
\nn
\eeqar
in the
corresponding representation $\Rep=V,\Phi,\Psi$.
For singlet-, doublet- and triplet-SU(2) representations we have 
$\Tr_\Rep=0,1/2$ and $2$, respectively.
For $\Rep=\Phi$, since we parametrize 
the Higgs 
doublet in terms of its four physical 
components, we have
\beqar\label{defDynkinb}
\Tr_\Phi\left(
I^{\bar{a}}
I^{{b}}
\right )
=
\frac{1}{2}\sum_{\Phi_i,\Phi_j=H,\chi,\phi^\pm}
I^{\bar a}_{\Phi_i\Phi_j}
I^b_{\Phi_j\Phi_i}
.
\nn
\eeqar}
\beqar\label{betacoeffexp1}
e^2\betacoeff{V,a b}^{(1)}&=&
\frac{11}{3}
e^2 \,\Tr_V\left(
I^{\bar{a}}
I^{{b}}
\right )
=\frac{11}{3}\gw^2 \CA
\de^{\groupb}_{{a} b}
,\nl
e^2
\betacoeff{\Phi,a b}^{(1)}&=&
-\frac{1}{3}
e^2 \,\Tr_\Phi\left(
I^{\bar{a}}
I^{{b}}
\right )
=-\frac{1}{6}
\left[\gb^2 
Y_{\Phi}^2
\de^{\groupa}_{{a} b}
+\gw^2 
\de^{\groupb}_{{a} b}
\right],
\nl
e^2
\betacoeff{\Psi,a b}^{(1)}&=&
-\frac{2}{3}
e^2\sum_{\rho=\pm}
\Tr_\Psi\left(
I^{\bar{a}}_\rho
I^{{b}}_\rho
\right )
=-\frac{2}{3}
\left[
\gb^2 
\sum_{\rho=\pm}
\frac{\left(Y_{\rho}\right)_d^2
+\left(Y_{\rho}\right)_u^2
}{4}
\,\de^{\groupa}_{{a} b}
+\frac{1}{2}\gw^2 
\de^{\groupb}_{{a} b}
\right]
,
\eeqar
where the last line corresponds to
a doublet of leptons or quarks
($\Psi=\mathrm{L,Q}$) with up- and down- components $\Psi_i=u,d$. 
The resulting contributions to \refeq{betamatrixel}
read 
\beqar\label{betacoeffexpe1}
e^2 \betacoeff{V,e}^{(1)}&=&
\frac{11}{3}
e^2\,\Tr_V\left(
I^{{A}}
I^{{A}}
\right )
=\frac{11}{3}\gw^2\sw^2 \CA
,\nl
e^2
\betacoeff{\Phi,e}^{(1)}&=&
-\frac{1}{3}
e^2 \,\Tr_\Phi\left(
I^{{A}}
I^{{A}}
\right )
=-\frac{1}{6}
\left[
\gb^2\cw^2Y_\Phi^2+\gw^2\sw^2 
\right],
\nl
\betacoeff{\Psi,e}^{(1)}&=&
-\frac{4}{3}
\Tr_\Psi\left(
I^{{A}}
I^{{A}}
\right )
=-\frac{4}{3}
\sum_{\Psi_i=u,d}
Q_{\Psi_i}^2
.
\eeqar
The 
QED $\beta$-function coefficient is given by the light-fermion contributions only,
\ie
\beqar\label{betacoeff2}
\betacoeff{\QED}^{(1)}
&=&
\betacoeff{F,e}^{(1)}
+\frac{4}{3}N_c
Q_{\Pt}^2.
\eeqar

\end{appendix}

\end{document}